%% file: sinaioscarxiv.tex
\documentclass[amsmath,amssymb,floatfix,prl]{revtex4}

\pdfoutput=1

\usepackage[pdftex]{graphicx}
\usepackage{amssymb,amsfonts,amsmath}
\usepackage{graphicx,amsmath}
\usepackage{float}

\DeclareMathOperator{\sgn}{sgn}

\def\Aberg{{\AA}berg\ }
\def\eps{\varepsilon}
\def\pdag{{\phantom\dagger}}
\def\aa{(\textbf{a})}
\def\bb{(\textbf{b})}
\def\cc{(\textbf{c})}
\def\dd{(\textbf{d})}
\def\ee{(\textbf{e})}
\def\ff{(\textbf{f})}

\begin{document}

\title{Dynamical thermalization of interacting fermionic atoms in a Sinai-oscillator trap}

\author{Klaus~M.~Frahm}
\affiliation{\mbox{Laboratoire de Physique Th\'eorique, IRSAMC, 
Universit\'e de Toulouse, CNRS, UPS, 31062 Toulouse, France}}
\author{Leonardo Ermann}
\affiliation{\mbox{ Departamento de F\'\i sica Te\'orica, GIyA, Comisi\'on Nacional 
de Energ\'ia At\'omica, Buenos Aires, Argentina}}
\author{Dima L. Shepelyansky}
\affiliation{\mbox{Laboratoire de Physique Th\'eorique, 
Universit\'e de Toulouse, CNRS, UPS, 31062 Toulouse, France}}

\date{July 16, 2019}

\begin{abstract}
We study numerically 
the problem of dynamical thermalization of interacting cold fermionic atoms
placed in an isolated Sinai-oscillator trap. 
This system is characterized by a quantum chaos regime for one-particle dynamics.
We show that for a many-body system of cold atoms the interactions, with a strength above 
a certain quantum chaos border given by the \Aberg criterion,
lead to the Fermi-Dirac distribution 
and relaxation of many-body initial states
to the thermalized state in absence of any contact with a thermostate.
We discuss the properties of this dynamical thermalization and its links
with the Loschmidt-Boltzmann dispute.
\end{abstract}

\maketitle

\subsection{I Introduction} 
\label{sec1}

The problem of emergence of thermalization in
dynamical systems started from the Loschmidt-Boltzmann dispute
about time reversibility and thermalization 
in an isolated system of moving and colliding classical atoms \cite{loschmidt,boltzmann}
(see the modern overview in \cite{mayer,jalabert}). 
The modern resolution of this dispute is related
to the phenomenon of dynamical chaos where an exponential instability of
motion breaks the time reversibility at infinitely small 
perturbation (see e.g.  \cite{arnold,sinaibook,chirikov,lichtenberg}).
The well known example of such a chaotic system
is the Sinai billiard in which 
a particle moves inside a square box with
an internal circle  colliding elastically with 
all boundaries \cite{sinaibil}. 

The properties of one-particle quantum systems, which are chaotic in the classical limit,
have been extensively studied in the field of quantum chaos 
during the last decades and their properties have
been mainly understood (see e.g.  \cite{gutzwiller,haake,stockmann}).
Thus it was shown that the level spacing statistics in the 
regime of quantum chaos \cite{bohigas1984} is the same as for 
Random Matrix Theory (RMT) 
invented by Wigner for a description of spectra of complex nuclei
\cite{wigner,mehta}. 
This result became known as the Bohigas-Giannoni-Schmit conjecture \cite{bohigas1984,ullmo}.
Thus classically chaotic systems (e.g. Sinai billiard) are usually characterized
by Wigner-Dyson (RMT) statistics with level repulsion \cite{wigner,mehta,bohigas1984} 
while the classically integrable systems  usually show Poisson statistics
for level spacing distribution \cite{haake,stockmann,ullmo}. In this way the level 
spacing statistics gives a direct indication for ergodicity (Wigner-Dyson statistics)
or non-ergodicity (Poisson statistics) of quantum eigenstates. It was also established that
the classical chaotic diffusion can be suppressed by quantum interference effects leading to
an exponential localization of eigenstates 
\cite{chirikov1981,fishman,chirikov1988,fishmanschol}
being similar to the Anderson localization in disordered solid-state systems \cite{anderson}.  
The localized phase is characterized by Poisson statistics 
and the delocalized or metallic phase has RMT statistics. For billiard systems
the localized (nonergodic) and delocalized (ergodic)  regimes appear in the case of 
rough billiards as described in \cite{rough1,rough2}.

It was also shown that in the regime of quantum chaos  
the Bohr correspondence principle \cite{bohrcorres} and the
fully correct semiclassical description of quantum evolution
remain valid only for a logarithmically short Ehrenfest time scale 
$t_E \sim \ln(1/\hbar)/h$ \cite{chirikov1981,chirikov1988}. 
Here $\hbar$ is an effective dimensionless Planck constant and 
$h$ in the Kolmogorov-Sinai entropy characterizing the exponential 
divergence of classical trajectories.
This result is in agreement with the Ehrenfest theorem,
which states that the classical-quantum correspondence works on a time scale 
during which the wave packet remains compact \cite{ehrenfest}. However, for the 
classically chaotic systems the Ehrenfest time scale is rather short
due to an exponential instability of classical trajectories.
After the Ehrenfest  time scale $t_E$
the quantum out-of-time correlations (or OTOC as it is used to say now)
stop to decay exponentially in contrast to exponentially decaying 
classical correlators \cite{dls1981,dls1983}. 
For $t > t_E$ the decay of quantum correlations 
stops and they remain on the level of quantum fluctuations
being proportional to $\hbar$  \cite{dls1981,dls1983,stmapscholar}.
Since the level of quantum fluctuations is proportional to $\hbar$
the classical diffusive spreading over the momentum
is affected by quantum corrections only on a significantly larger
diffusive time scale $t_D \propto 1/\hbar^2 \gg t_E \propto \ln(1/\hbar)$
\cite{chirikov1981,chirikov1988,dls1981,dls1983,stmapscholar}.

The problem of emergence of RMT statistics and quantum ergodicity in many-body quantum systems
is more complex and intricate as compared to one-particle quantum chaos.
Indeed, it is well known that in many-body quantum systems 
the level spacing between nearest energy levels drops exponentially with the increase of 
number of particles or with energy excitation $\delta E$ above the Fermi level
in finite size Fermi systems, e.g. in nuclei \cite{bohr}.  
Thus on a first glance it seems that an exponentially small
interaction between fermions should mix many-body quantum levels
leading to RMT level spacing statistics (see e.g. \cite{guhr}). 

Furthermore, the size of the Hamiltonian matrix of a many-body system
grows exponentially with the number of particles
but since all interactions have a two-body nature the number of nonzero interaction
elements in this matrix grows not faster than 
the number of particles in fourth power.
Thus we have a very sparse matrix being rather far from the RMT 
type. A two-body random interaction model (TBRIM) 
was proposed in \cite{french1,bohigas1} 
to consider the case of generic random two-body interactions of fermions
in the limiting case of strong interactions when  
one-particle orbital energies are neglected.
Even if the TBRIM matrix is very sparse, 
is was shown that the level spacing statistics $p(s)$ 
is described by 
the Wigner-Dyson or RMT distribution \cite{french2,bohigas2}.

However, it is also important to analyze another limiting case
when the two-body interaction matrix elements of strength 
$U$ are weak or comparable with 
one-particle energies with an average level spacing $\Delta_1$.
In metallic quantum dots this case with $U/\Delta_1 \approx 1/g$
corresponds to a large conductance of a dot
$g = E_{Th}/\Delta_1 \gg 1$ where $E_{Th} =\hbar/t_D$ is the Thouless energy
with $t_D$ being a diffusion spread  time 
over the dot \cite{thouless,imry,akkermans}. 
In this case the main question
is about critical interaction strength $U$ or excitation energy
$\delta E$ above the Fermi level of the dot
at which the RMT statistics becomes valid. First  numerical results and 
simple estimates for a critical interaction strength in
a model similar to TBRIM were 
obtained by Sven {\AA}berg in \cite{aberg1,aberg2}.
The estimate of a critical interaction $U_c$, 
called the {\AA}berg criterion \cite{dlsnobel},
compares the typical two-body matrix elements
with the energy level spacing $\Delta_c$ 
between quantum states {\it directly coupled by 
two-body interactions}. Thus  the {\AA}berg criterion tells that
the Poisson statistics is valid for many-body energy levels
for $U < U_c \sim \Delta_c$ and the RMT statistics
sets in for $U > U_c \sim \Delta_c$. 
In \cite{jacquod} this criterion, proposed 
independently of \cite{aberg1,aberg2},
was applied to the TBRIM of weakly interacting fermions in a metallic quantum dot
being confirmed by extensive numerical simulations.
It was also argued that the dynamical thermalization
sets in an isolated finite fermionic system 
for energy excitations $\delta E$ above the critical border $\delta E_{ch}$ 
determined from the above criterion \cite{jacquod}:
\begin{equation}
\label{eq_g23}
\delta E > \delta E_{ch} \approx g^{2/3} \Delta_1 \; , \;\; g = \Delta_1/U \; .
\end{equation}
The emergence of thermalization in an isolated many-body system
induced by interactions between particles without any contact with an external thermostat
represents the Dynamical Thermalization Conjecture (DTC) proposed in \cite{jacquod}.
The validity of the  {\AA}berg criterion was numerically confirmed for
various physical models (see \cite{dlsnobel} and Refs. therein). 
An additional confirmation was given
by  the analytical derivation presented in \cite{sushkov}
showing that for 3 interacting particles in a metallic dot 
the RMT sets in when the two-body matrix elements $U$
become larger than the two-particle level spacing
$\Delta_2 \sim \Delta_c $ being parametrically larger
than the three-particle level spacing $\Delta_3 \ll \Delta_2$. The advanced
theoretical arguments developed in \cite{mirlin1,mirlin2}
confirm the relation (\ref{eq_g23}) for interacting fermions 
in a metallic quantum dot.

The test for the transition from Poisson to RMT statistics is 
rather direct and needs only the knowledge of energies eigenvalues. 
However, the verification of DTC is much more involved
since it requires the computation of system eigenstates.
Thus it is much more difficult to check numerically the relation (\ref{eq_g23})
for DTC. However, it is possible to show that there a transition
from non-thermalized eigenstates at weak interactions 
(presumably for $\delta E < \delta E_{ch}$)
to dynamically thermalized individual eigenstates
at relatively strong interactions 
(presumably for $\delta E > \delta E_{ch}$).
Thus for the TBRIM with fermions
the validity of DTC for individual eigenstates
at $U> U_c \sim \Delta_c$
has been demonstrated in \cite{kolovsky2017,frahmtbrim}
by the computation of energy $E$ and entropy $S$
of each eigenstate and its comparison
with the theoretical result given
by the Fermi-Dirac thermal distribution \cite{landau}.

Even if the TBRIM represents a useful system for
DTC tests it is not so easy to realize it 
in real experiments. Thus, in this work we
investigate the DTC features in a system of
cold fermionic atoms
placed in the Sinai-oscillator trap created by a harmonic two-dimensional
potential with a repulsive circular potential created by a laser beam
in a vicinity of the trap center. 
In such a case the repulsive potential in the center
is modeled as an elastic circle as in the case of Sinai billiard \cite{sinaibil}.
For one particle it has been shown in \cite{ermannsinai}
that the Sinai oscillator has an almost fully chaotic phase space
and that in the quantum case the level spacing statistics is described
by the RMT distribution. Due to one-particle quantum chaos in the Sinai oscillator
we expect that this system will have properties similar of the TBRIM.
On the other side the Sinai-oscillator trap has been already 
experimentally realized  with 
Bose-Einstein condensate of cold bosonic atoms
\cite{ketterle1,ketterle2,ketterle3}. At present cold atom techniques
allow to investigate various properties of cold 
interacting fermionic atoms \cite{roati1,roati2} and we argue that 
the investigation of dynamical thermalization of such
fermionic atoms, e.g. $^6Li$, in a Sinai-oscillator trap
is now experimentally possible. 
Thus in this work we study properties of DTC
of interacting fermionic atoms in a Sinai-oscillator trap.
Here, we consider the two-dimensional (2D) case of such
a system assuming that the trap frequency 
in the third direction is small and that the 
2D dynamics is not significantly affected
by the adiabatically slow motion in the third dimension. 

Finally, we note that
at present the TBRIM model in the limit of strong interactions
attracts a high interest in the context of field theory 
since in this limit it can be mapped on a black hole model  
of quantum gravity in $1 + 1$ dimensions known as the Sachdev-Ye-Kitaev 
(SYK) model linked also to a strange metal 
\cite{sachdevprl,kitaev,sachdevprx,rosenhaus,maldacena,garcia1}.
In fact, the SYK model, in its fermionic formulation \cite{sachdevprx},  
corresponds  to  the  TBRIM  considered  with a conductance
close  to  zero $g \rightarrow 0$. In these lines the dynamical thermalization
in TBRIM and SYK systems has been discussed in \cite{kolovsky2017,frahmtbrim}.
Furthermore, there is also a growing interest in dynamical thermalization
for various many-body systems known also as the eigenstate thermalization 
hypothesis (ETH) and many-body localization (MBL) 
(see e.g. \cite{huse,polkovnikov,borgonovi,alet}).
We think that the system of interacting fermionic atoms in a Sinai-oscillator 
trap captures certain features of TBRIM and SYK models 
and thus represents an interesting test ground to 
investigate nontrivial physics of these systems in real cold atom experiments.

This paper is composed as follows: in Section 2 we describe the properties
of the one-particle dynamics in a Sinai oscillator; 
numerical results for dynamical thermalization
on interacting atoms  in this oscillator are presented in Section 3;
the conditions of thermalization for fermionic cold atoms
in realistic experiments are given in Section 4;
the discussion of the results is presented in Section 5.

\section{II Quantum chaos in Sinai oscillator}
\label{sec2}

The model of one particle in the 2D Sinai oscillator is described in 
detail in \cite{ermannsinai} with the Hamiltonian: 
\begin{equation}
H_1=\frac{1}{2m}(p_x^2+p_y^2)+
\frac{m}{2}(\omega_x^2 x^2+\omega_y^2 y^2)+V_d(x,y) \;\; .
\label{eq_sinaiham}
\end{equation}
Here the first two terms describe the 2D oscillator with 
frequencies $\omega_x, \omega_y$ and the last term gives
the potential wall of elastic disk of radius  $r_d$.
We choose the dimensionless units with mass $m=1$, 
frequencies $\omega_x=1$, $\omega_y=\sqrt{2}$
and disk radius $r_d=1$. The disk center is located at
 $(x_d,y_d)=(-1/2,-1/2)$ so that the disk bungs a hole in the center
as it was the case in the experiments \cite{ketterle1}.
The Poincare sections at different energies are presented in \cite{ermannsinai}
showing that the phase space is almost fully chaotic (see Figure 1 there).
The quantum evolution is described by the Schr\"odinger equation
with the quantized Hamiltonian (\ref{eq_sinaiham}) 
where the conjugate momentum and coordinate variables become
operators with the commutation relation $[x,p_x] = [y,p_y] = i\hbar$ \cite{ermannsinai}.
For the quantum problem we use the value of the dimensionless Planck constant
$\hbar=1$ so that the ground state energy is $E_g = 1.685$.
In the following the energies are expressed in atomic like units 
of energy $E_u = \hbar \omega_x$ (for our choice of
Sinai oscillator parameters we also have 
$E_u = \hbar \omega_x = \hbar^2/(m{r_d}^2)$) \cite{ermannsinai}
with the typical size of oscillator ground state being equal to the disk 
radius:
$a_0 = \Delta x_{osc} = (\hbar/m \omega_x)^{1/2} = r_d$.

\begin{figure}[H]
\centering
\includegraphics[width=0.8\textwidth]{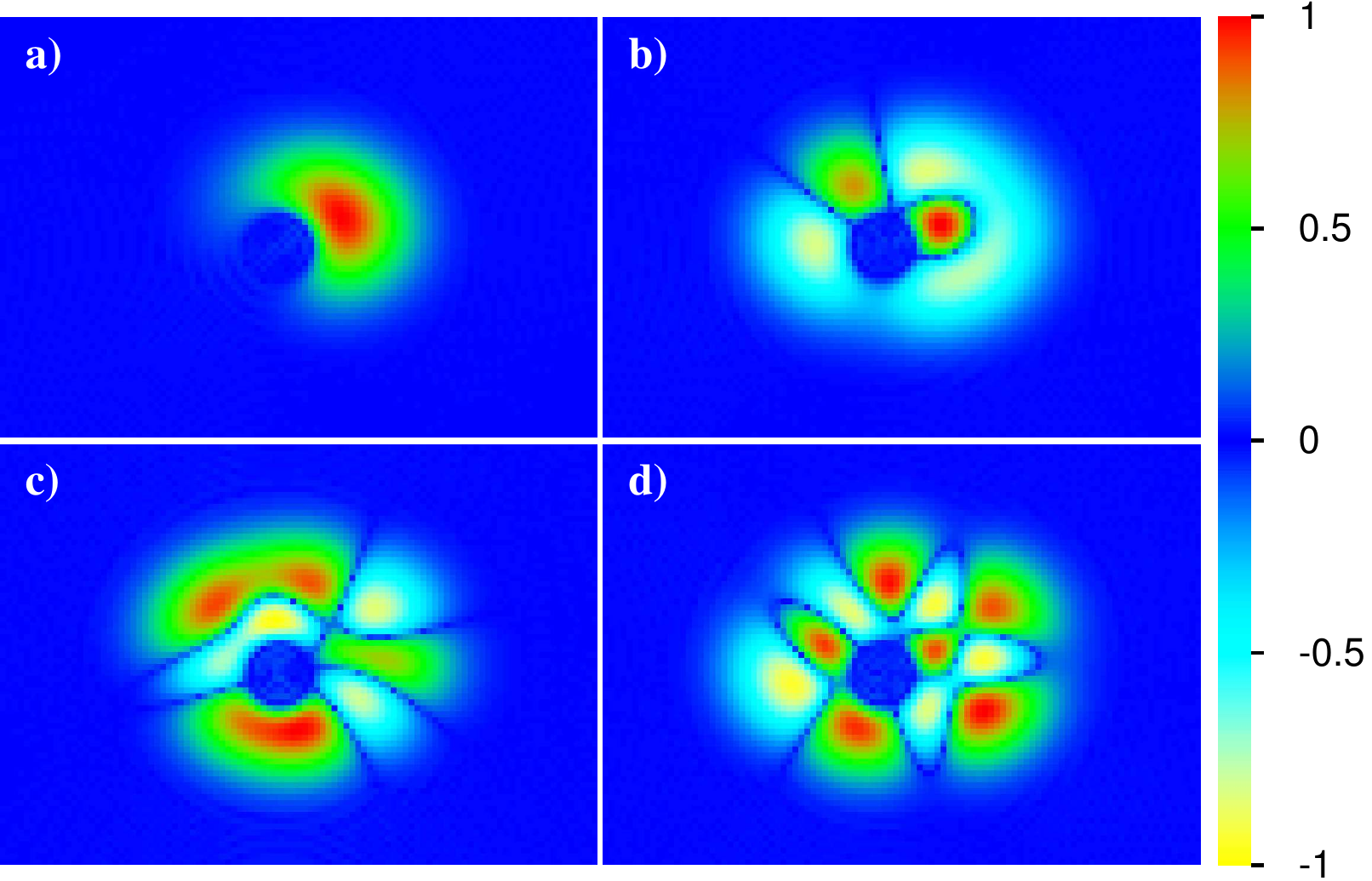}
\caption{\label{fig1} 
Color plot of one-particle eigenstates $\varphi_k(x,y)$ of the 
Sinai Hamiltonian in coordinate plane $(x,y)$ with 
$-7.6\le x\le 7.6$ and $-5.4\le y\le 5.4$ for orbital numbers 
$k=1$ (ground state) \aa , $k=6$ \bb , $k=11$ \cc\ and $k=16$ \dd. 
The numerical values of the color bar apply to the signed and 
nonlinearly rescaled wave function amplitude:  
$\sgn[\varphi_k(x,y)]\,|\varphi_k(x,y)/\varphi_{\rm max}|^{1/2}$ 
where $\varphi_{\rm max}$ is the maximum of $|\varphi_k(x,y)|$ and 
the exponent $1/2$ provides amplification of regions of small amplitude.
}
\end{figure}

In \cite{ermannsinai} it is shown that the classical dynamics of this system
is almost fully chaotic. In the quantum case the level spacing statistics
is well described by the RMT distribution.
The average dependence of energy level number $k$ is 
well described by the theoretical dependence 
$k(\eps) =  \eps^2/(2\sqrt{2}) - \eps/2$ \cite{ermannsinai}.
Thus the one-particle density of states $\rho_1(\eps)$ and 
corresponding level spacing $\Delta_1$ are:
\begin{equation}
\label{eq_1pDOS}
\rho_1(\eps) =\frac{dk}{d\eps} = \frac{\eps}{\sqrt{2}}-\frac{1}{2}
\quad,\quad\Delta_1 =\frac{1}{\rho} \approx \frac{\sqrt{2}}{\eps} \approx 
\frac{0.84}{\sqrt{k}}.
\end{equation}
Examples of several eigenstates, computed on a numerical grid 
of 28341 spatial points, are shown in Figure~\ref{fig1}.
More details on the numerical diagonalization of (\ref{eq_sinaiham}) 
and other example eigenstates can be found in \cite{ermannsinai}.

\section{III Sinai oscillator with interacting fermionic atoms}
\label{sec3}

\subsection{Two-body interactions of fermionic atoms}

The two-body interaction of atoms appears usually due to 
van der Waals forces which drop rapidly with the distance between two atoms 
and the short ranged interaction can be described in the frame work of 
the scattering length approach (see e.g. \cite{gribakin1,gribakin2}).
Therfore we assume that the finite effective interaction range $r_c$ is 
significantly smaller than the disk radius $r_d$ and the typical size of
the wave function, i.e. $r_c \ll r_d$. Such a short range interaction is 
indeed used 
to modelize atomic interactions in harmonic traps (see e.g. \cite{wilkens}). 
For example in a typical experimental situation the disk radius is of the 
order of micron 
$r_d \sim 1 \mu m = 10^{-4} cm$ while for Li and other alkali atoms
we have $r_c \sim 3 \times 10^{-7} cm$ \cite{gribakin1,gribakin2}.

In the following, we use a simple interaction function having a 
constant amplitude $U$ for $r\le r_c$ and being zero for $r>r_c$ where 
we simply choose $r_c=0.2 r_d$ which corresponds well to the short range 
interaction regime. 
The precise value of $r_c$ is not very important since a slight modification 
$r_c\to \bar r_c$ can be absorbed in a modified amplitude according to 
$U\to \bar U=U(\bar r_c/r_c)^2$, a relation we verified numerically 
for various values of $\bar r_c<r_d$. 
We mention that in experiments the strength of the interaction amplitude 
can be changed by a variation of the magnetic field via the 
Feshbach resonance \cite{kohler}.

\subsection{Reduction to TBRIM like case and its analysis}

Using the methods described in \cite{ermannsinai}
we numerically compute a certain number of one-particle or orbital 
energy eigenvalues 
$\eps_k$ and corresponding eigenstates $\varphi_k({\bf r})$ 
of the Sinai oscillator (\ref{eq_sinaiham}). As repulsive 
interaction potential $v({\bf r})$ 
we choose the short ranged box function $v({\bf r})=U$ if 
$|{\bf r}|\le r_c=0.2$ (since $r_d=1$) and $v({\bf r})=0$ otherwise. 
Here the parameter 
$U>0$ gives the overall scale of the interaction strength depending on the 
charge of the particles and eventually other physical parameters. 

Therefore the corresponding many-body Hamiltonian 
with $M$ one-particle orbitals and $0\le L\le M$ spinless fermions
takes the form:
\begin{equation}
\label{eq_HamMB}
H=\sum_{k=1}^M \eps_k\,c^\dagger_k c^\pdag_k
+\sum_{i<j,k<l} V_{ij,kl}\,c^\dagger_i c^\dagger_j 
c^\pdag_l c^\pdag_k
\end{equation}
where for $i<j$ and $k<l$ we have the interaction matrix elements:
\begin{equation}
\label{eq_Vdef}
V_{ij,kl}=\bar V_{ij,kl} -\bar V_{ij,lk}\quad,\quad
\bar V_{ij,kl}=\int d{\bf r_1}\int d{\bf r_2}\ \varphi_i^*({\bf r_1}) 
\varphi_j^*({\bf r_2})\,v({\bf r_1}-{\bf r_2}) \varphi_k({\bf r_1}) 
\varphi_l({\bf r_2})
\end{equation}
and $c^\dagger_k$, $c^\pdag_k$ are fermion operators for the $M$ orbitals 
satisfying the usual anticommutation relations. We note that in the literature, 
when expressing a two-body interaction potential in second quantization, 
one usually uses the raw matrix elements $\bar V_{ij,kl}$ with an 
additional prefactor of $1/2$ and 
full independent sums for the four indices $i$, $j$, $k$ and $l$. 
Using the particle exchange symmetry: $\bar V_{ij,kl}=\bar V_{ji,lk}$
one can reduce the $i,j$ sums to $i<j$ which removes the prefactor 
$1/2$ (after exchanging the index names $l$ and $k$ for the $i>j$ 
contributions and 
exploiting that contributions at $i=j$ or $l=k$ obviously vanish). The 
definition of the anti-symmetrized interaction matrix elements 
$V_{ij,kl}$ according 
to (\ref{eq_Vdef}) allows to reduce also the $k,l$ sums to $k<l$. 
Furthermore, the ordering of the two fermion operators $c^\pdag_l c^\pdag_k$ 
in (\ref{eq_HamMB}) is also important and necessary to obtain positive 
expectation values if the interaction is repulsive. 
The anti-symmetrized matrix elements $V_{ij,kl}$ correspond to a 
$M_2\times M_2$ matrix with $M_2=M(M-1)/2)$. In order to avoid a global 
shift of the non-interacting eigenvalue spectrum due to the interaction
we also apply a diagonal shift 
$V_{ij,ij}\to V_{ij,ij}-(1/M_2)\sum_{k<l} V_{kl,kl}$ to ensure that 
this matrix has a vanishing trace\footnote{One can easily show that 
the trace of the $M_2\times M_2$ anti-symmetrized interaction matrix 
is proportional to the trace of the interaction operator in the many-body 
Hilbert space with a factor depending on $M$ and $L$.}. Of course 
such a global energy shift and does not affect the issues of thermalization, 
interaction induced eigenfunction mixing or the quantum time evolution 
with respect to the Hamiltonian $H$ etc. 

\subsection{\Aberg parameter}

In absence of interaction the energy eigenvalues of (\ref{eq_HamMB}) 
are given as the sum of occupied orbital energies:
\begin{equation}
\label{eq_E1pdef}
E(\{n_k\})=\sum_{k=1}^M \eps_k\,n_k
\end{equation}
where $\{n_k\}$ represents a configuration such that $n_k\in\{0,1\}$ 
and $\sum_k n_k=L$. The associated eigenstates are the basis states 
where each orbital is either occupied (if $n_k=1$) or unoccupied (if 
$n_k=0$) and in this work we will denote these states in the usual 
occupation number representation: $|n_M\cdots n_2\,n_1\!\!>$ 
where for convenience we write the lower index orbitals starting 
from the right side. 

The distribution of the total one-particle energies 
(\ref{eq_E1pdef}) is numerically rather close to a Gaussian (since $n_k$ 
act as quasi-random numbers) with mean and variance (see also Eq. (A.4) of 
Ref. \cite{frahmtbrim}):
\begin{equation}
\label{eq_Emean_var}
E_{\rm mean}=L\overline\eps\quad,\quad
\sigma_0^2=\frac{L(M-L)}{M-1}\,
\left(\overline{\eps^2}-\overline\eps^2\right)\quad,\quad 
\overline{\eps^n}=\frac{1}{M}\sum_{k=1}^M\eps_k^n\quad,\quad n=1,2.
\end{equation}
Therefore the many-body level spacing $\Delta_{\rm MB}$ or inverse 
Heisenberg time at the band center $E=E_{\rm mean}$ is given by 
$\Delta_{\rm MB}=1/t_{\rm H}=\sqrt{2\pi}(\sigma_0/d)$ where 
$d=M!/(L!(M-L)!)$ is the dimension of the fermion Hilbert space 
in the sector of $M$ orbitals and $L$ particles. In our numerical 
computations we simply evaluated the quantities $\overline{\eps^n}$ 
of (\ref{eq_Emean_var}) using the exact one-particle energy eigenvalues 
obtained from the numerical diagonalization 
of the one-particle Sinai Hamiltonian $H_1$ given in (\ref{eq_sinaiham}). 
However, to get some analytical simplification for large $M$ one 
may use the one-particle 
density of states (\ref{eq_1pDOS}) which gives, after replacing 
the sums by integrals and neglecting the constant term, 
$\overline{\eps^n}\approx 2\,\eps_M^n/(n+2)$ and 
$\overline{\eps^2}-\overline{\eps}^2\approx \eps_M^2/18\approx \sqrt{2}\,M/9$. 

For the question if the interaction strength is sufficiently strong to 
mix the non-interacting basis states the important quantity is the effective 
level spacing of states coupled directly by the interaction 
$\Delta_c=\sqrt{2\pi}\,[\sigma_0(L=2)/K]$ where 
$K=1 + L(M-L) + L(L-1)(M-L)(M-L-1)/4$ is the number 
of nonzero elements for a column (or row) of $H$ \cite{jacquod,flambaum} 
and we need to use the variance for only two particles:
\begin{equation}
\label{eq_sigam2p}
\sigma_0^2(L=2)=\frac{2(M-2)}{M-1}\,
\left(\overline{\eps^2}-\overline\eps^2\right)\quad\Rightarrow\quad
\frac{\sigma_0^2(L=2)}{\sigma_0^2}=\frac{2(M-2)}{L(M-L)}
\end{equation}
because the interaction only couples states where (at least) $L-2$ particles 
are on the same orbital such that (at most) only the partial sum of two 
one-particle energies is different between two coupled states. Even though 
for two particles the hypothesis of a Gaussian distribution is theoretically 
not justified the distribution is still sufficiently similar to a Gaussian 
and it turns out 
that the value of $1/\Delta_c=K/[\sqrt{2\pi}\,\sigma_0(L=2)]$ as 
the coupled 
two-particle density of states in the band center is numerically quite 
accurate with an error below 10 \% (for $M=16$ and our choice of $\eps_k$ 
values). 

According to the \Aberg criterion \cite{aberg1,aberg2,jacquod} the onset 
of chaotic mixing happens for typical interaction matrix elements $U$ 
comparable to $\Delta_c$. Therefore we compute the quantity 
$V_{\rm mean}=\sqrt{\langle |V_{ij,kl}|^2\rangle}$ (which 
is proportional to the interaction amplitude $U$) 
where the average is done with respect to all $M_2^2$ matrix elements 
of the interaction matrix. This quantity might be problematic and 
not correspond to a typical interaction matrix element in case of a 
long tail distribution. However, in our case it turns out that 
$V_{\rm mean}\approx 2\exp(\langle \ln |V_{ij,kl}|\rangle)$ which 
excludes this scenario. Using this quantity we introduce the 
dimensionless \Aberg parameter and the critical interaction amplitude $U_c$ 
by 
$A=V_{\rm mean}/\Delta_c=U/U_c$ such 
that $A=1$ if $U=U_c$. 
We expect \cite{aberg1,aberg2,jacquod} the onset of strong/chaotic mixing 
at $A\gg 1$ and a perturbative regime for $A\ll 1$ while at $A=1$ we have 
the critical interaction strength $U=U_c$. 
The value of $U_c$ depends on the parameters 
$L$, $M$, $\sigma_0$ and the overlap of the 
one-particle eigenstates according to (\ref{eq_Vdef}). 
To obtain some useful analytical expression of $U_c$ we note that 
the quantity $V_{\rm mean}$, numerically 
computed for $4\le M\le 30$, can be quite 
accurately fitted by $V_{\rm mean}\approx 3\times 10^{-4}\,U/\eps_M$. 
Furthermore, we remind the expression 
$\Delta_c=(1/K)\,\sqrt{4\pi(M-2)(\overline{\eps^2}-\overline{\eps}^2)/(M-1)}$
which can be simplified in the limit $M\gg 1$ and $L\gg 1$, such 
that $K\approx (M-L)^2 L^2/4$, resulting in:
$\Delta_c=4/3\,\sqrt{2\pi}\,\eps_M/[(M-L)^2\,L^2]$. Here we also 
used the above found expression 
$\overline{\eps^2}-\overline{\eps}^2\approx \eps_M^2/18$. From this we 
find that $U_c=\Delta_c U/V_{\rm mean}\approx C\,M/[(M-L)^2\,L^2]$ 
with a numerical constant 
$C\approx 16\times 10^4\sqrt{\pi}/9\approx 3.15\times 
10^4$ where we also used $\eps_M^2\approx 2\sqrt{2}\,M$ 
according to (\ref{eq_1pDOS}). Below we will give more 
accurate numerical values of $V_{\rm mean}$, $\Delta_c$ and $U_c$ for 
the parameter choice of $M$ and $L$ numerically relevant in this work. 

We note that this estimate for $A=U/U_c$ applies to energies close to the 
many body band center of $H$ and that for energies away from the band 
center the 
value of $\Delta_c$ is enhanced thus reducing the effective value of $A$. 
Furthermore, we computed $V_{\rm mean}$ by a simplified average 
over {\em all} interacting matrix elements not taking into account an eventual 
energy dependence according to the index values of $i,j,k,l$ in 
(\ref{eq_Vdef}). 


\subsection{Density of states}

In this work we present numerical results for 
the case of $M=16$ orbitals and $L=7$ particles corresponding to a 
many-body Hilbert space of dimension $d=M!/(L!(M-L)!)=11440$ and 
the number $K=820$ of directly coupled states of a given initial state 
by non-vanishing interaction matrix elements in (\ref{eq_HamMB}). 
Thus in our studies the whole Hilbert space is built only on these 
$M=16$ orbitals.
We diagonalize numerically the many-body Hamiltonian (\ref{eq_HamMB}) 
for various values of $A$ in the range $0.025\le A\le 200$. 
We have also verified that the results and their physical interpretation 
are similar for smaller cases such as $M=12$, $L=5$ 
(with $d=792$, $K=246$) or $M=14$, $L=6$ ($d=3003$, $K=469$). 

We mention that for $M=16$ and $L=7$ we find numerically that 
$V_{\rm mean}=3.865\times 10^{-5}\,U$ and from 
(\ref{eq_sigam2p}) that 
$\Delta_c=\sqrt{2\pi}\,[\sigma_0(L=2)/K]=6.1706\times 10^{-3}$ where 
the quantities $\overline{\eps^n}$ where exactly computed
from the numerical orbitals energies $\eps_k$. From this we find that 
$U_c=\Delta_c\,U/V_{\rm mean}\approx 159.65$. This expression is 
more accurate than the more general analytical estimate for arbitrary 
$M\gg 1$ and $L\gg 1$ given in the last section (which would provide 
$U_c\approx 127$ for $M=16$ and $L=7$).

Our first observation is that, even in presence of interactions,
the density of states has approximately a Gaussian 
form with the same center $E_{\rm mean}$ given in (\ref{eq_Emean_var}) 
for the case $A=0$. This is simply due the fact that the 
interaction matrix has, by choice, a vanishing trace and does not provide a 
global shift of the spectrum. 
We determine the variance $\sigma^2(A)$ of the Gaussian density of states 
by a fit of the integrated density of states $P(E)$ using 
\begin{equation}
\label{eq_int_dos}
P(E)=(1+\mbox{erf}[q(E)])/2\quad,\quad q(E)=(E-E_{\rm mean})/
[\sqrt{2}\,\sigma(A)]
\end{equation}
such that $P'(E)$ is a Gaussian of width $\sigma(A)$ and center $E_{\rm mean}$ 
(see Appendix A of Ref. \cite{frahmtbrim} for more details). 
From this we find the behavior~:
\begin{equation}
\label{eq_sigma_int}
\sigma^2(A)=\sigma_0^2\,(1+\alpha A^2)
\end{equation}
where $\alpha$ is a constant depending on $M$ and $L$; for $M=16$, $L=7$ 
the fit values of $\sigma_0$ and $\alpha$ are $\sigma_0=3.013\pm 0.009$ and 
$\alpha=0.00877\pm 0.00010$. It is also 
possible to determine $\sigma(A)$ using the expression
$\sigma^2(A)=\mbox{Tr}_{\rm Fock}\left[(H-E_{\rm mean}{\bf 1})^2\right]/d$ 
where the trace in Fock space can be evaluated 
either by using the matrix $H$ before diagonalizing it or using its 
exact energy eigenvalues $E_m$. This provides the same behavior as 
(\ref{eq_sigma_int}) with the very similar numerical values 
$\sigma_0=3.013\pm 0.007$ and $\alpha=0.00858\pm 0.00008$ (for $M=16$, $L=7$). 
We mention that the integrated Gaussian density of states (\ref{eq_int_dos}) 
is not absolutely exact but quite accurate for values $A\le 10$. For larger 
values of $A$ the deviations increase but the overall form is still correct. 
As described in \cite{frahmtbrim} the quality of the fit can be considerably 
improved if we replace in (\ref{eq_int_dos}) the linear function $q(E)$ by 
a polynomial of degree $5$. In this case the precision of the fit is highly 
accurate for the full range of $A$ values we consider. In particular, 
we use this improved fit to perform the spectral unfolding when computing 
the nearest level spacing distribution (shown below).

To obtain some theoretical understanding of (\ref{eq_sigma_int}) one 
can consider a model where the initial interaction matrix elements 
(\ref{eq_Vdef}) are replaced by independent Gaussian variables with
 identical variance $V_{\rm mean}^2$. In this case one can show 
theoretically \cite{frahmtbrim}
that $\sigma^2(A)=\sigma_0^2+K_2 V_{\rm mean}^2$ where 
$K_2=L(L-1)[1 + M-L + (M-L)(M-L-1)/4]$ is a number somewhat larger 
than $K$ taking into account that certain non-vanishing 
interaction matrix elements in Fock space are given as a sum of 
{\em several} initial interaction matrix elements (\ref{eq_Vdef}) 
(see Appendix A of \cite{frahmtbrim} 
for details). The parameter $K_2$ takes for $M=16$, $L=7$ ($M=14$, $L=6$ 
or $M=12$, $L=5$) the value $K_2=1176$ ($K_2=690$ or $K_2=370$ respectively). 
Since $V_{\rm mean}=A\Delta_c=A\sqrt{2\pi}\,\sigma_0(L=2)/K$ 
we indeed obtain (\ref{eq_sigma_int}) 
with $\alpha=\alpha_{\rm th}=4\pi (M-2)K_2/[K^2(L(M-L)]$. 
For $M=16$, $L=7$ we find $\sigma_0=3.0279$ (see (\ref{eq_Emean_var})) 
and $\alpha_{\rm th}=0.00488$. The latter 
is roughly by a factor of $2$ smaller than the numerical value. We 
attribute this to the fact that the real initial interaction matrix 
elements (\ref{eq_Vdef}) are quite correlated, and not independent 
uniform Gaussian variables, leading therefore to an effective increase of 
the number $K_2$ due to hidden correlations. 
The important point is that theoretically at very large values values of 
$M$ and $L$, e.g. $M\approx 2L\gg 1$ we have 
$K_2\approx K\approx L^4/4$ and $\alpha_{\rm th}\approx 32\pi/L^5$ 
which is parametrically small for very large $L$. Therefore, there 
is a considerable range of values $1<A<1/\sqrt{\alpha}$ where the 
interaction strongly mixes the non-interacting many-body eigenstates 
but where the density of states is only weakly affected by the interaction. 
This regime is also known as the Breit-Wigner regime 
(see e.g. \cite{dlsnobel} for the case of interacting Fermi systems). 

\subsection{Thermalization and entropy of eigenstates}

In the following, we mostly concentrate on values $A\le 10$ such that the 
effect of the increase of the spectral width $\sigma(A)$ is still small 
or at least quite moderate. The question arises if a given many-body state, 
either an exact eigenstate of $H$ or a state 
obtained from a time evolution with respect to $H$, is thermalized according 
to the Fermi-Dirac distribution \cite{landau}. 
As in \cite{kolovsky2017,frahmtbrim} we determine 
the occupation numbers $n_k=\langle c^\dagger_k c^\pdag_k\rangle$ for 
such a state, as well as the corresponding fermion entropy $S$ \cite{landau}
and the effective total one-particle energy $E_{1p}$ by~:
\begin{equation}
\label{eq_entropy}
S=-\sum_{k=1}^M\Bigl(n_k\,\ln n_k+(1-n_k)\,\ln(1-n_k)\Bigr)
\quad,\quad
E_{1p}=\sum_{k=1}^M\,\eps_k\,n_k
\end{equation}
based on the assumption of weakly interacting fermions. 
In the regime of modest interaction $A\lesssim 5$ 
(for $M=16$, $L=7$), corresponding to 
a constant spectral width $\sigma(A)\approx \sigma_0$, we have typically 
$E_{1p}\approx E_{\rm ex}$ (for exact eigenstates of $H$) 
or $E_{1p}\approx \langle H\rangle$ (for other states). If the given state 
is thermalized its occupation numbers $n_k$ should be close to 
the theoretical Fermi-Dirac filling factor 
$n(\eps_k)$ with $n(\eps)=1/(1+\exp[\beta(\eps-\mu)])$ where 
inverse temperature $\beta = 1/T$ and chemical potential $\mu$ are determined 
by the conditions:
\begin{equation}
\label{eq_beta_mu_cond}
L=\sum_{k=1}^M n(\eps_k)\quad,\quad 
E=\sum_{k=1}^M \eps_k\,n(\eps_k). 
\end{equation}
Here $E$ is normally given by $E_{1p}$ but one may also consider the 
value $E_{\rm ex}$ (or $\langle H\rangle$) provided the latter is in 
the energy interval where the conditions (\ref{eq_beta_mu_cond}) allow for a 
unique solution. Furthermore, for a given energy $E$ we can also 
determine the theoretical (or thermalized) entropy $S_{\rm th}(E)$ 
using (\ref{eq_entropy}) with $n_k$ being replaced by $n(\eps_k)$ (where 
$\beta$, $\mu$ are determined from (\ref{eq_beta_mu_cond}) for the 
energy $E$). 

\begin{figure}[H]
\centering
\includegraphics[width=0.8\textwidth]{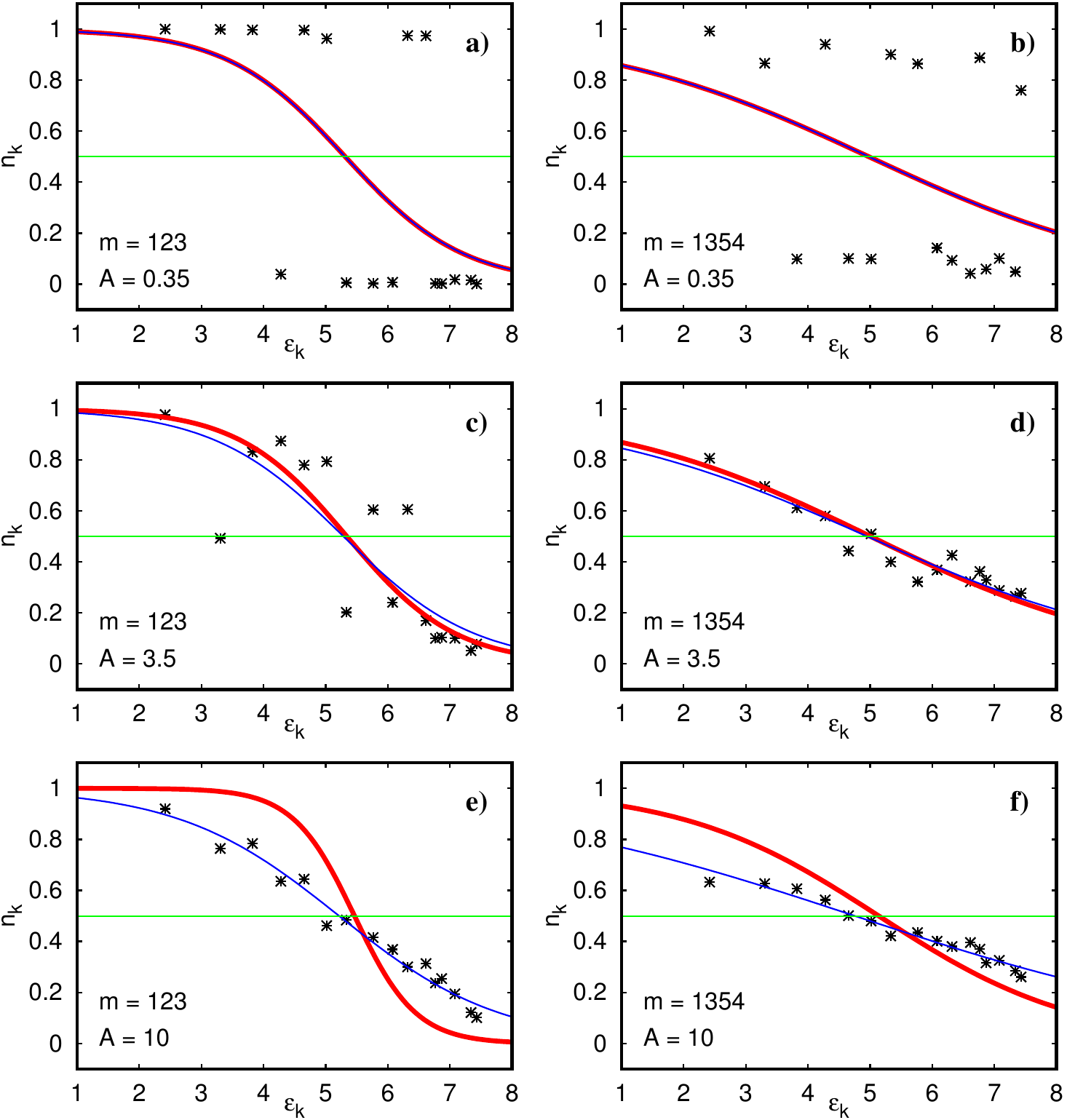}
\caption{\label{fig2} Orbital occupation number $n_k$ versus 
orbital energies $\eps_k$ (black stars) of individual eigenstates at 
level numbers $m=123$ \aa, \cc, \ee, $1354$ \bb, \dd, \ff\ 
and \Aberg parameter 
$A=0.35$ \aa, \bb, $A=3.5$ \cc, \dd, $A=10$ \ee, \ff\ (with $m=1$ 
corresponding to the ground state). 
The thin blue (thick red) curves show the 
theoretical Fermi-Dirac occupation number 
$n(\eps)=1/(1+\exp[\beta(\eps-\mu)])$ 
where inverse temperature $\beta$ and chemical potential 
$\mu$ are determined from (\ref{eq_beta_mu_cond}) 
with $E=E_{\rm 1p}$ ($E=E_{\rm ex}$). 
The horizontal green lines correspond to the constant value $0.5$ 
whose intersections with the red or blue curves provide the positions 
of the chemical potential. 
In this and all subsequent figures the orbital number is $M=16$,
the number of particles is $L=7$ and the corresponding 
dimension of the many body Hilbert space is $d=11440$. 
Table~\ref{table1} gives for each of these levels the values 
of $E_{\rm ex}$, $E_{\rm 1p}$, $S$, $S_{\rm th}$, $\beta$, $\mu$ 
and for both energies for the latter three parameters.}
\end{figure}

\begin{table}[H]
\caption{\label{table1} Parameters of the eigenstates corresponding 
to Figure~\ref{fig2}. $S$ is the entropy, $E_{1p}$ the 
effective total one-particle energy, both given by (\ref{eq_entropy}), 
and $E_{\rm ex}$ is the exact energy eigenvalue. 
Inverse temperature $\beta$, chemical potential 
$\mu$, theoretical entropy $S_{\rm th}$  
are determined by (\ref{eq_beta_mu_cond}) 
or (\ref{eq_entropy}) (with $n_k$ replaced by $n(\eps_k)$) 
for both energies $E_{1p}$, $E_{\rm ex}$.}
\centering
\include{table1pr}
\end{table}

The many-body states with energies above $E_{\rm mean}$ are artificial 
since they correspond to negative temperatures due to the finite number of 
orbitals considered in our model. Therefore we  limit our studies 
to the lower half of the energy spectrum $29\le E\le 39\approx E_{\rm mean}$ 
(for $M=16$, $L=7$). In Figure~\ref{fig2} we compare the thermalized 
Fermi-Dirac occupation number $n(\eps)$ with the the occupation numbers 
$n_k$ for two eigenstates at level numbers $m=123$ ($1354$; with $m=1$ 
corresponding to the ground state) with approximate energy eigenvalue 
$E\approx 32$ ($E\approx 35$) for three different \Aberg parameters 
$A=0.35$, $A=3.5$ and $A=10$. These states are not too close to the 
ground state but still quite far below the band center. 

At weak interaction, $A=0.35$, both states are not at all thermalized 
with occupation numbers being either close to 
1 or 0. Apparently these states result from weak perturbations of the 
non-interacting eigenstates $|0000011000110111\!\!>$ or 
$|1000100011001011\!\!>$ where the $n_k$ values are rounded to 1 (or 0) 
if $n_k>0.5$ ($n_k<0.5$). For $m=1354$ the values of $n_k$ are a little 
bit farther away from the ideal values $0$ or $1$ as compared to 
$m=123$ but still sufficiently close to be considered as perturbative. 
Apparently the state $m=123$, which is lower in the spectrum (with 
larger effective two-body level spacing), is less affected 
by the interaction than the state $1354$. In both cases the entropy 
$S$ is quite below the thermalized entropy $S_{\rm th}$ (see 
Table~\ref{table1} for numerical values of entropies, energies, 
inverse temperature and chemical potential for the states 
shown in Figure~\ref{fig2}). 

At intermediate interaction, $A=3.5$, the occupation numbers are closer to the 
theoretical Fermi-Dirac values but still with considerable deviations. 
Here both entropy values $S$ are rather close to $S_{\rm th}$. The 
state $1354$ seems to be better thermalized than 
the state $m=123$, the latter having a slightly larger deviation between 
both entropy values. 
At stronger interaction, $A=10$, both states are very well thermalized with 
a good matching of both entropy values (again with the state $1354$ 
being a bit better thermalized than the state $m=123$) 
provided we use $E_{\rm 1p}$ as reference energy 
to compute temperature and chemical potential. The temperature obtained 
from $E_{\rm ex}$ is too small because here the increase of $\sigma(A)$ 
is already quite strong and $E_{\rm ex}$ rather strongly deviates 
from $E_{\rm 1p}$. Also the value of $S_{\rm th}$ using $E_{\rm ex}$ does not 
match $S$. Obviously at stronger interaction values it is necessary 
to use $E_{\rm 1p}$ to test the thermalization hypothesis of a given state. 

\begin{figure}[H]
\centering
\includegraphics[width=0.8\textwidth]{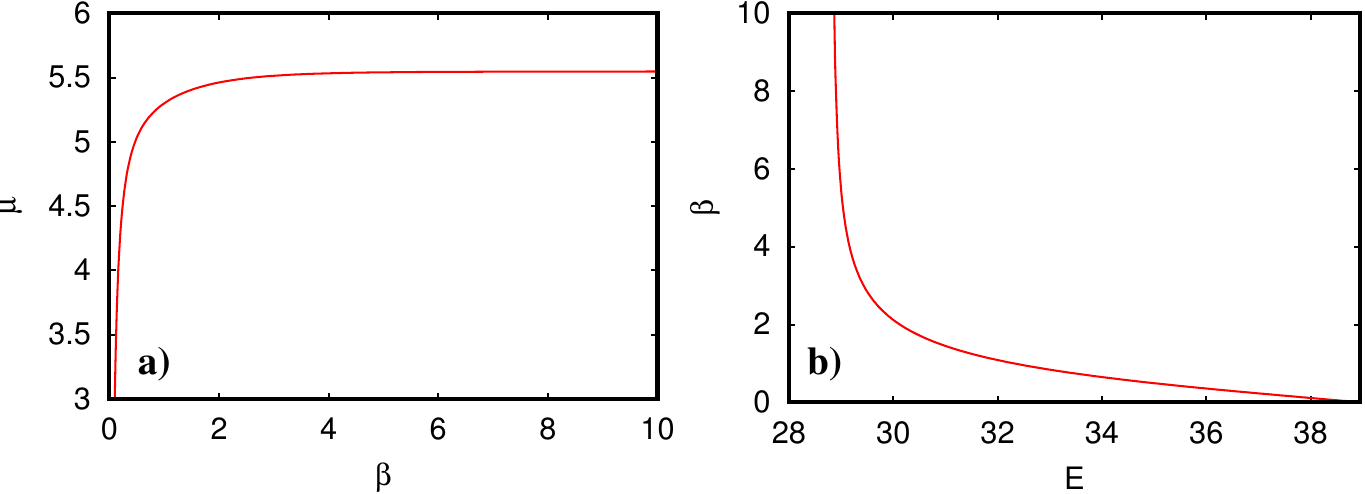}
\caption{\label{fig3} 
Dependence of chemical potential $\mu$ on inverse temperature $\beta=1/T$ 
\aa\ and of $\beta =1/T$ on energy $E$ \bb\ where $\beta$ and $\mu$ are 
determined from (\ref{eq_beta_mu_cond}) for a given energy $E$. 
}
\end{figure}

Figure~\ref{fig3} shows the mutual dependence between the three 
parameters $\beta$, $\mu$ on $E$ when solving the conditions 
(\ref{eq_beta_mu_cond}). The chemical potential as a function of $\beta =1/T$ 
is rather constant except for smallest values of $\beta$ where 
$\mu \sim 1/\beta$ with a negative prefactor. One can actually 
easily show from (\ref{eq_beta_mu_cond}) that in the limit $\beta\to 0$ 
the chemical potential does not depend on $\eps_k$ and is given by 
$\mu=-\,\ln[1+(M-2L)/L]/\beta$ providing a singularity if $L\neq M/2$ 
with negative (positive) prefactor for $L<M/2$ ($L>M/2$) and $\mu=0$ 
for $L=M/2$. 
The temperature ($\beta^{-1}$) 
vanishes for $E$ close to the lower energy border 
and diverges for $E$ close to the band center $E_{\rm mean}$. 

\begin{figure}[H]
\centering
\includegraphics[width=0.8\textwidth]{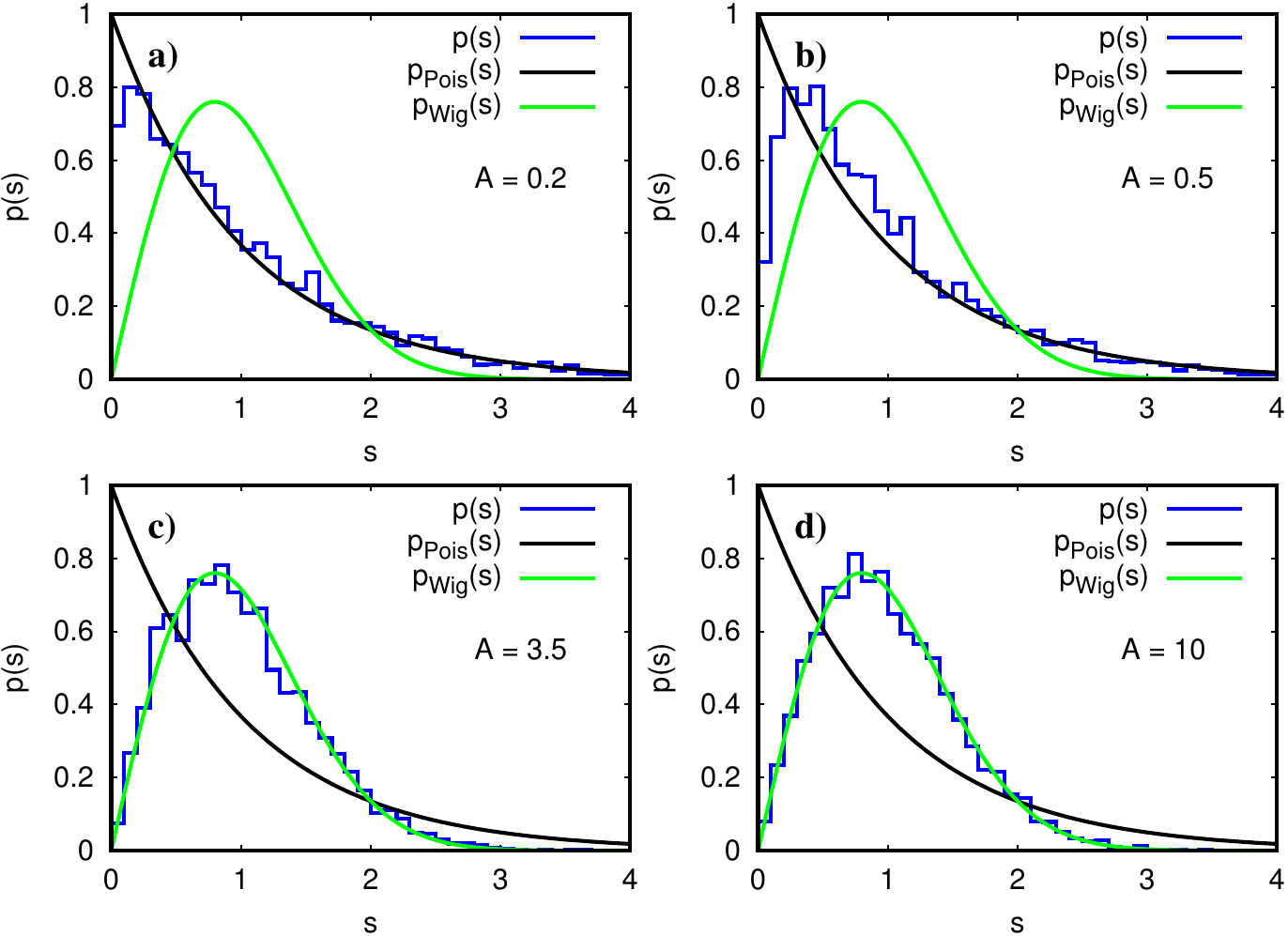}
\caption{\label{fig4} 
Histogram of unfolded level spacing statistics (blue line) 
for the exact energy eigenvalues $E_m$ of $H$ (using 
the lower half of the spectrum with $1\le m \le d/2$). 
The different panels correspond to the \Aberg parameter values 
$A=0.2$ \aa, $A=0.5$ \bb, $A=3.5$ \cc, $A=10$ \dd. 
The unfolding is done using the integrated density of states 
(\ref{eq_int_dos}) where $q(E)$ is replaced by a fit polynomial of degree $5$. 
The Poisson distribution $p_{\rm Pois}(s)=\exp(-s)$ (black line) 
and the Wigner surmise 
$p_{\rm Wig}(s)=\frac{\pi}{2}\,s\,\exp(-\frac{\pi}{4}\,s^2)$ 
(green line) are also shown for comparison.
}
\end{figure}

In Figure~\ref{fig4} we present 
the nearest level spacing distribution $p(s)$ for 
different values of the \Aberg parameter. To compute $p(s)$ we have used 
only the ``physical'' levels in the lower half of the energy spectrum and the 
unfolding has been done with the integrated density of states 
(\ref{eq_int_dos}) where $q(E)$ is replaced by a fit polynomial of 
degree 5. For the smallest value $A=0.2$ the distribution 
$p(s)$ is very close to the Poisson distribution with some 
residual level repulsion at very small spacings. This is a quite 
well known effect because typically the transition from Wigner-Dyson to 
Poisson statistics (when tuning some suitable parameter such as the 
\Aberg parameter from strong to weak coupling) 
is non-uniform in energy and happens first at 
larger spacings (energy differences) and then at smaller spacings. 
The reason is simply that two levels which by chance are initially 
very close are easily repelled by a small residual coupling matrix element 
(when slightly changing a disorder realization or similar). 
For $A=0.5$ there is somewhat more level repulsion at small spacings but 
the distribution is still rather close to the Poisson distribution with 
some modest deviations for $s\le 1.2$. 
For the larger \Aberg values $A=3.5$ and $A=10$ we clearly obtain Wigner-Dyson 
statistics (taking into account the quite limited number of 
only $d/2-1=5719$ level spacing values for the histograms). 
These results clearly confirm that the transition from 
$A<1$ to $A>1$ corresponds indeed to a transition from a perturbative 
regime to a regime of chaotic mixing with Wigner-Dyson level 
statistics \cite{wigner}. 

\begin{figure}[H]
\centering
\includegraphics[width=0.8\textwidth]{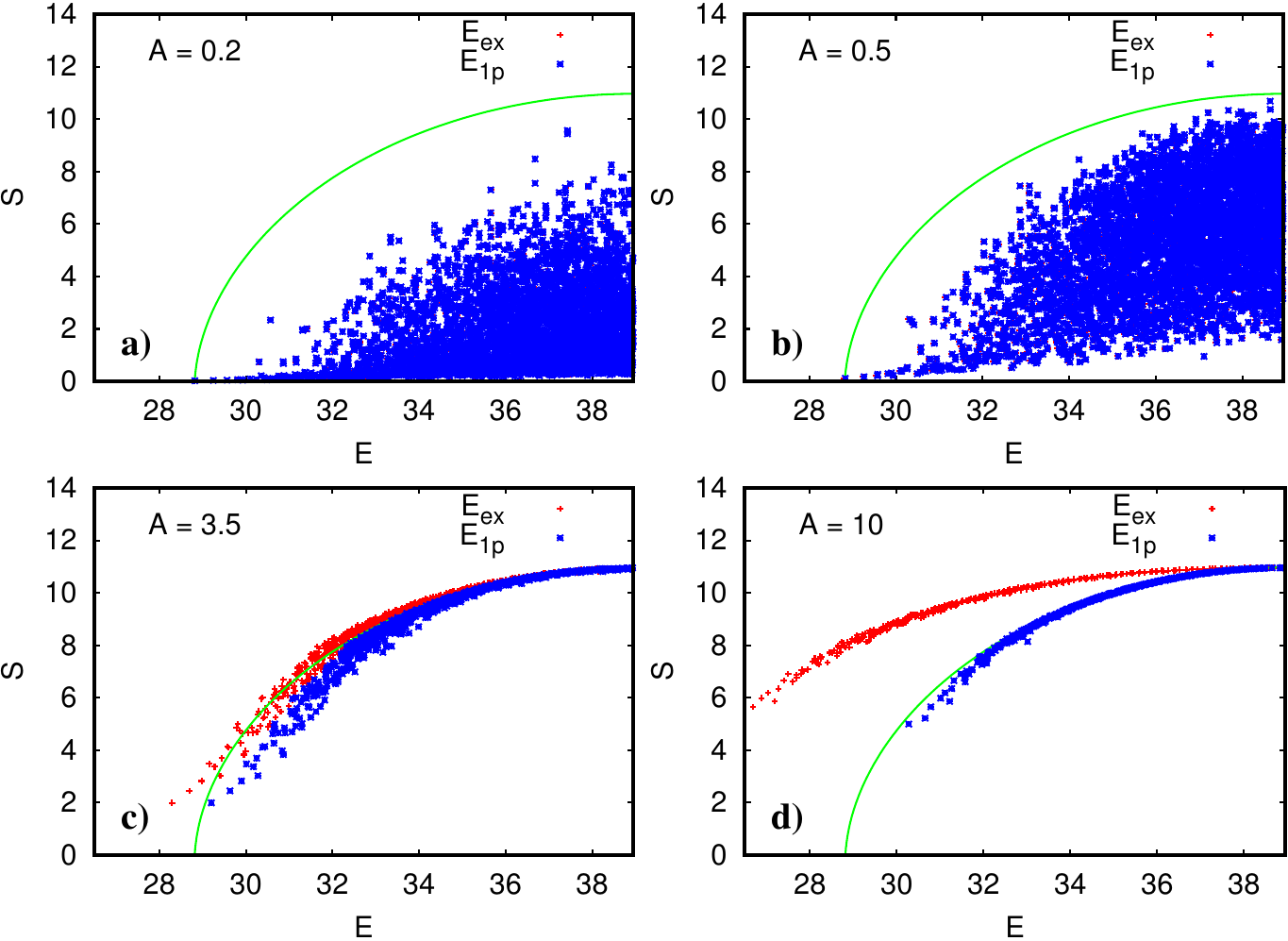}
\caption{\label{fig5} Dependence of the fermion entropy $S$ 
on the effective one-particle total energy 
$E_{1p}$ (blue cross symbols) 
and the exact many-body energy $E_{\rm ex}$ (red plus symbols). 
The green curve shows the theoretical entropy $S_{\rm th}(E)$ 
obtained from the Fermi-Dirac occupation numbers as explained in the text. 
The different panels correspond to the \Aberg parameter values 
$A=0.2$ \aa, $A=0.5$ \bb, $A=3.5$ \cc, $A=10$ \dd. 
}
\end{figure}

A further confirmation that $A=1$ is critical can be seen 
in Figure~\ref{fig5} which compares 
the dependence of the entropy $S$ of exact eigenstates (lower half of 
the spectrum) on $E_{\rm 1p}$ or $E_{\rm ex}$ with the theoretical 
thermalized entropy $S_{\rm th}(E)$. For the \Aberg values $A=0.2$ (and 
$A=0.5$) the entropy $S$ of all (most) states is significantly below its 
theoretical value $S_{\rm th}$. Actually the distribution of data points 
is considerably concentrated at smaller entropy values which is not 
so clearly visible in the Figure. In particular 
the average of the ratio of $S/S_{\rm th}(E_{\rm 1p})$ is 
$0.178$ for $A=0.2$ and $0.522$ for $A=0.5$. 
For the \Aberg values $A=3.5$ and $A=10$ most or nearly all entropy values 
(for $E_{\rm 1p}$) are very close to the theoretical line 
with the average ratio $S/S_{\rm th}(E_{\rm 1p})$ being $0.990$ 
for $A=3.5$ and $0.998$ for $A=10$. For $A=3.5$ the states with lowest 
energies are not yet perfectly thermalized and the data points 
for $E_{\rm ex}$ and $E_{\rm 1p}$ are still rather close. 
For $A=10$ all states are well thermalized (when using the 
energy $E_{\rm 1p}$) 
while the data points for $E_{\rm ex}$ are quite outside the theoretical curve 
simply due to the overall increase of the width of the energy spectrum. This 
observation is also in agreement with the discussion of Figure~\ref{fig2}. 
For smaller values $A<0.2$ (not shown in Figure~\ref{fig5}) we find that 
the data points are still closer to the $E$-axis while 
for larger values $A>10$ the data points are clearly on  
the theoretical curve for $E_{\rm 1p}$ (but more concentrated on 
energy values closer to the center with larger entropy values and larger 
temperatures) while for $E_{\rm ex}$, according to (\ref{eq_sigma_int}), 
the overall width of the exact eigenvalue spectrum increases strongly 
and the data points are clearly outside the theoretical curve (except 
for a few states close to the band center).

\begin{figure}[H]
\centering
\includegraphics[width=0.8\textwidth]{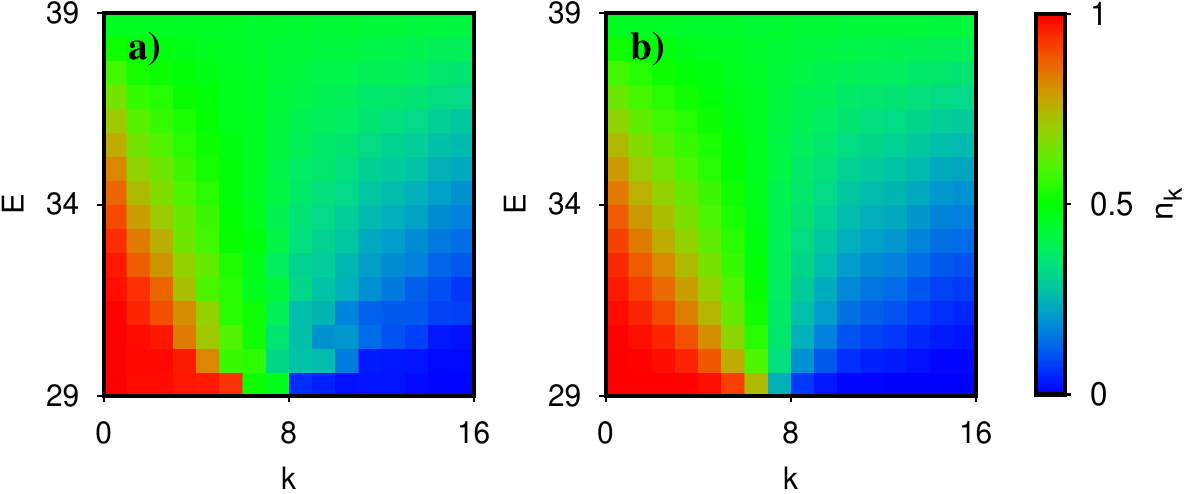}
\caption{\label{fig6} 
Color density plot of the orbital occupation number $n_k$ in the plane 
of energy $E$ and orbital index $k$. \aa\ $n_k$ values of exact 
eigenstates of $H$ with \Aberg parameter 
$A=3.5$; \bb\ thermalized 
Fermi-Dirac occupation number $n(\eps_k)$ where $\beta$ and $\mu$ are 
determined from (\ref{eq_beta_mu_cond}) as a function of total energy $E$.
The occupation number $n_k$ is averaged over all 
eigenstates \aa\ or several representative values of $E$ \bb\ inside 
a given energy cell. The energy interval $29\le E\le 39$ corresponds 
roughly to the lower half of the spectrum (at $M=16$, $L=7$) for states 
with positive temperature and is similar to the energy interval used 
in Figures~\ref{fig4} and \ref{fig5}. 
The color bar provides the translation between $n_k$ 
values and colors (red for maximum $n_k=1$, green for $n_k=0.5$ and blue 
for minimum $n_k=0$). 
}
\end{figure}

In Figure~\ref{fig6} the occupation numbers $n_k$ (averaged over 
several energy eigenvalues inside a given energy cell) are shown 
in the plane of energy $E$ and orbital index $k$ as color density 
plot for the \Aberg parameter $A=3.5$. The comparison with the theoretical 
occupation numbers $n(\eps_k)$ (shown in the same way) 
provides further confirmation that at $A=3.5$ there is indeed already 
a quite strong thermalization of most eigenstates. 

\subsection{Thermalization of quantum time evolution}

The question arises how or if a time dependent 
state $|\psi(t)\!\!>=\exp(-iHt)|\psi(0)\!\!>$, 
obeying the quantum time evolution with the Hamiltonian $H$ 
and an initial state $|\psi(0)\!\!>$ being a non-interacting eigenstate 
$|n_M\cdots n_2 n_1)\!\!>$ (with all $n_k\in\{0,\,1\}$
and $\sum_k n_k=L$), evolves eventually 
into a thermalized state. We have computed such time dependent states 
using the exact eigenvalues and eigenvectors of $H$ to evaluate the 
time evolution operator. As initial states we have chosen four states 
(for $M=16$, $L=7$): (i) $|\phi_1\!\!>=|0000100000111111\!\!>$ 
where a particle at orbital 7 is excited from the non-interacting ground 
state (with all orbitals from 1 to 7 occupied) to the orbital 12, 
(ii) $|\phi_2\!\!>=|0010100000011111\!\!>$ 
where two particles at orbitals 6 and 7 are excited from the non-interacting 
ground state to the orbitals 12 and 14, (iii) 
$|\phi_3\!\!>=|0000011000110111\!\!>$ and (iv) 
$|\phi_4\!\!>=|1000100011001011\!\!>$. The states $|\phi_3\!\!>$ and 
$|\phi_4\!\!>$ are obtained from the exact eigenstate of $H$ for $A=0.35$ 
at level number $m=123$ and $1354$ respectively 
by rounding the occupation numbers $n_k$ to 1 (or 0) if $n_k>0.5$ 
($n_k<0.5$) (states of top panels in Figure~\ref{fig2}). The 
approximate energies (\ref{eq_E1pdef}) of these four states are 
$E\approx 30$ ($|\phi_1\!\!>$), $E\approx 32$ ($|\phi_2\!\!>$ and 
$|\phi_3\!\!>$) and $E\approx 35$ ($|\phi_4\!\!>$).

It is useful to express the time in multiples of the elementary quantum 
time step defined as:
\begin{equation}
\label{eq_timestep}
\Delta t=\frac{t_{\rm H}}{d}=\frac{1}{\sqrt{2\pi\,\sigma^2(A)}}
\end{equation}
where $t_{\rm H}$ is the Heisenberg time (at the given value of $A$), 
$d$ the dimension of the Hilbert space and $\sigma(A)$ the width 
of the Gaussian density of states given in (\ref{eq_sigma_int}). 
The quantity $\Delta t$ is the shortest physical time scale of the system (inverse 
of the largest energy scale) and obviously for $t\ll \Delta t$ the 
unitary evolution operator is close to the unit matrix multiplied by a uniform 
phase factor: 
$\exp(-iHt)\approx \exp(-iE_{\rm mean}t)\,{\bf 1}$ since the 
eigenvalues $E_m$ of $H$ satisfy $|E_m-E_{\rm mean}|\lesssim \sigma(A)$. 
We expect that any signification deviation of $|\psi(t)\!\!>$ with 
respect to the initial condition $|\psi(0)\!\!>$ happens at 
$t\ge\Delta t$ (or later in case of very weak interaction). Furthermore, 
by analyzing the time evolution in terms of the ratio $t/\Delta t$ 
the results do not depend on the global energy scale of the spectral width. 
The longest time scale is the Heisenberg time $t_{\rm H}\approx 10^4\Delta t$ 
(since $d=11440$ for $M=16$, $L=7$). Later we  also discuss 
intermediate time scales such as the inverse decay rate obtained 
from the Fermi golden rule. 

To show graphically the time evolution we compute the time dependent 
occupation numbers $n_k(t)=<\!\!\psi(t)|c^\dagger_k c^\pdag_k|\psi(t)\!\!>$
and present them in a color density plot in the plane $(k, \;t/\Delta t)$. 
Also, at the last used time value we compute the effective total 
one-particle energy $E_{\rm 1p}$ using the relation (\ref{eq_entropy}) 
(note that $E_{\rm 1p}$ is not conserved with respect to the time evolution 
except for very weak interaction) 
and use this value to determine from (\ref{eq_beta_mu_cond}) the inverse 
temperature $\beta$, chemical potential $\mu$ and the thermalized Fermi-Dirac 
filling factor $n(\eps_k)$ at each $k$ value for the orbital index. 
These values of ideally thermalized occupation numbers will be shown 
in an additional vertical bar \footnote{Note that this additional 
bar is not related to the usual color bar that provides the translation 
of colors to $n_k$ values. The latter is shown in Figure~\ref{fig6} and 
applies also to all subsequent figures with color density plots for $n_k$ 
values.} right behind the data for the last time values separated by a 
vertical white line. This presentation allows for an easy verification 
if the occupation numbers at the last time values are indeed thermalized 
or not. 

\begin{figure}[H]
\centering
\includegraphics[width=0.8\textwidth]{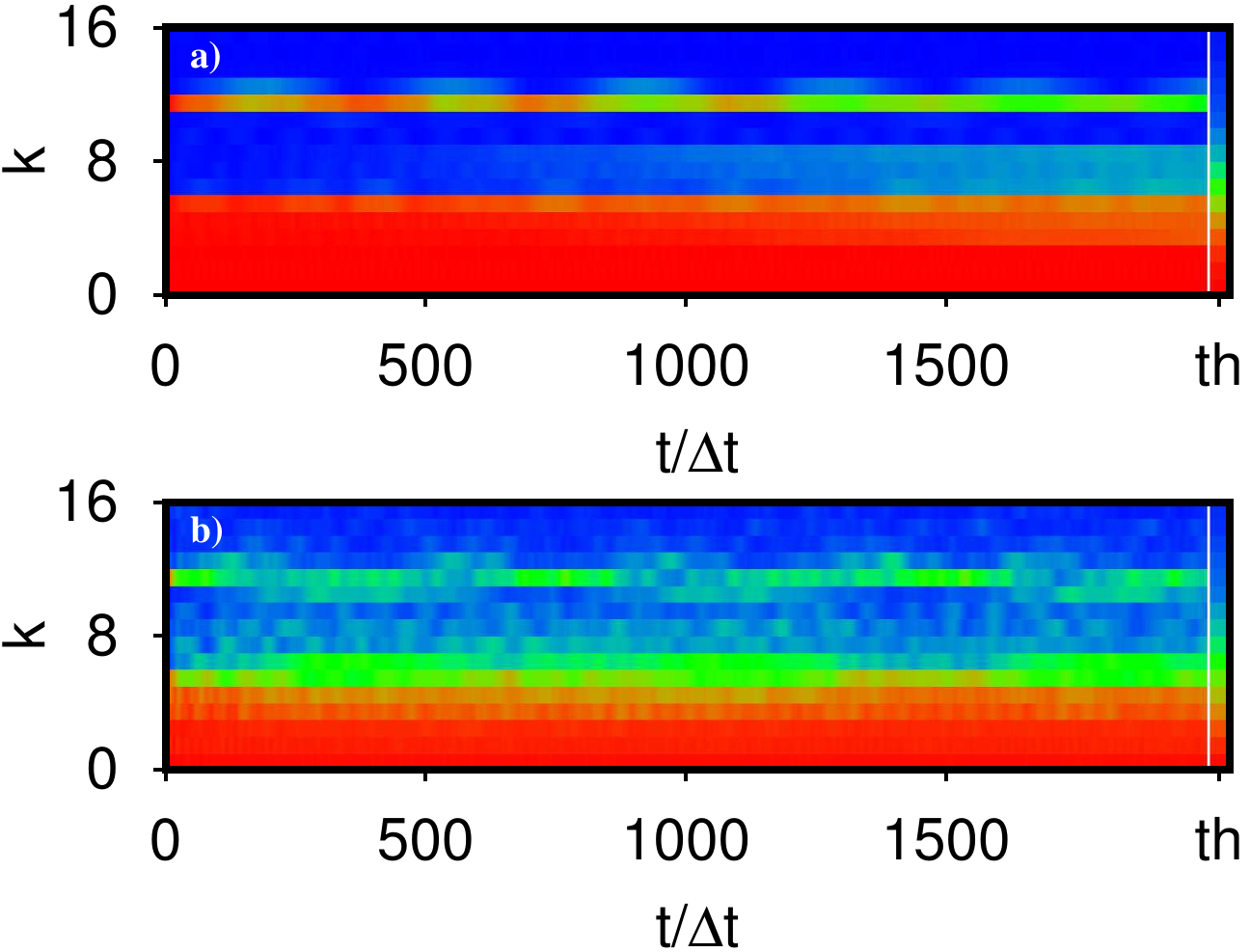}
\caption{\label{fig7} 
Color density plot of the orbital occupation number $n_k$ in the plane 
of orbital index $k$ and time $t$ for the time dependent state 
$|\psi(t)\!\!>=\exp(-iHt)|\psi(0)\!\!>$ with initial 
condition $|\psi(0)\!\!>=|\phi_1\!\!>=|0000100000111111\!\!>$. 
The time values are integer multiples of the elementary quantum time step 
$\Delta t=t_{\rm H}/d=1/[\sqrt{2\pi}\,\sigma(A)]$ where $t_{\rm H}$ is the 
Heisenberg time (at the given value of $A$). The bar behind the vertical 
white line with the label ``th'' shows the theoretical thermalized 
Fermi-Dirac occupation numbers $n(\eps_k)$ where $\beta$ and $\mu$ are 
determined from (\ref{eq_beta_mu_cond}) using the energy $E=E_{\rm 1p}$ 
of the state $|\psi(t)\!\!>$ at the last time value $t=2000\,\Delta t$. 
The two panels correspond to the \Aberg parameter $A=1$ \aa,
$A=3.5$ \bb.
For the translation of colors to $n_k$ values the color bar of 
Figure~\ref{fig6} applies. 
}
\end{figure}

In Figure~\ref{fig7} we show the time evolution for the initial 
state $|\phi_1\!\!>$ and the two \Aberg parameter values $A=1$ and 
$A=3.5$ using a linear time scale with integer multiples of $\Delta t$ 
and for $t\le 2000\,\Delta t\approx t_{\rm H}/6$. At $A=1$ the 
occupation number $n_{\rm 12}$ (of the excited particle) shows at the 
beginning a periodic structure, with an approximate period $400\,\Delta t$ 
for $t<1000\,\Delta t$, and a modest decay for $t>1000\,\Delta t$. 
At the same time the first orbitals above the Fermi sea are slightly excited. 
At final $t=2000\,\Delta t$ the state is clearly not thermalized. 
For $A=3.5$ we see a very rapid partial decay of $n_6$ and $n_{\rm 12}$ 
together with an increase of $n_{\rm 7}$. Furthermore for $n_k$ with 
$8\le k\le 11$ there are later and more modest excitations with a periodic 
time structure. Here the final state at $t=2000\,\Delta t$ is also not 
thermalized but it is closer to thermalization as for the case $A=1$. 

\begin{figure}[H]
\centering
\includegraphics[width=0.8\textwidth]{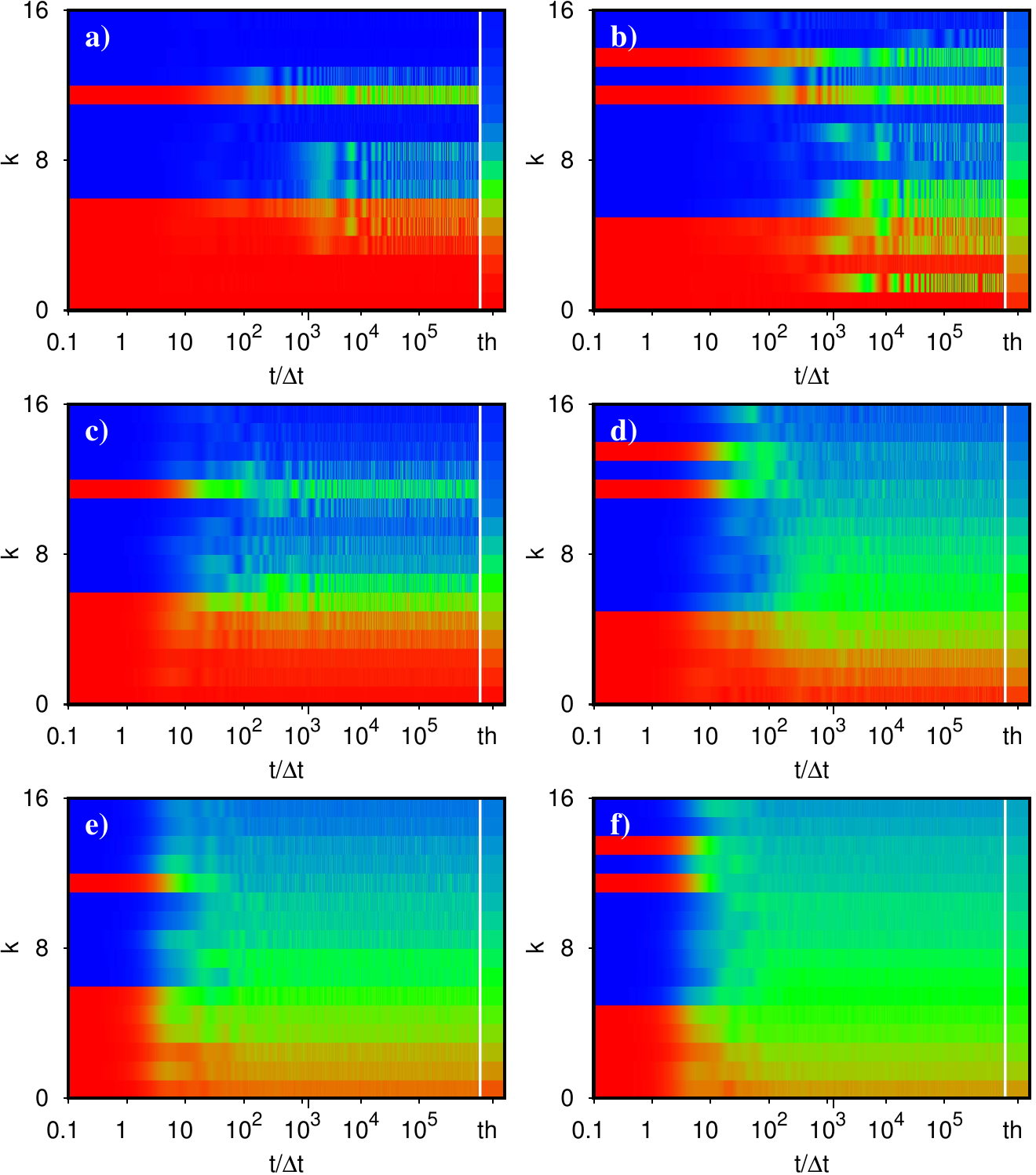}
\caption{\label{fig8} 
Color density plot of the orbital occupation number $n_k$ in the plane 
of orbital index $k$ and time $t$ for the time dependent state 
$|\psi(t)\!\!>=\exp(-iHt)|\psi(0)\!\!>$. 
The time axis is shown in logarithmic scale with time values 
$t_n=10^{\,(n/100)-1}\,\Delta t$ and integer $n\in\{0,1,\ldots,700\}$ 
corresponding to $0.1\le t_n/\Delta t\le 10^6$. The elementary quantum time 
step $\Delta t$ is the same as in Figure~\ref{fig7}. The bar behind the 
vertical white line with the label ``th'' shows the theoretical thermalized 
Fermi-Dirac occupation numbers $n(\eps_k)$ where $\beta$ and $\mu$ are 
determined from (\ref{eq_beta_mu_cond}) using the energy $E=E_{\rm 1p}$ 
of the state $|\psi(t)\!\!>$ at the last time value $t=10^6\,\Delta t$. 
The additional longer tick below the $t$-axis right next 
to the tick for $10^{3}$ gives the position of the maximal time value 
$t/\Delta t=2000$ of Figure~\ref{fig7}. The different panels 
correspond to the initial state 
$|\psi(0)\!\!>=|\phi_1\!\!>=|0000100000111111\!\!>$ \aa, \cc, \ee\ or 
$|\psi(0)\!\!>=|\phi_2\!\!>=|0010100000011111\!\!>$ \bb, \dd, \ff\ 
and \Aberg parameter values 
$A=1$ \aa, \bb, $A=3.5$ \cc, \dd, $A=10$ \ee, \ff.
For the translation of colors to $n_k$ values the color bar of 
Figure~\ref{fig6} applies. 
}
\end{figure}

\begin{figure}[H]
\centering
\includegraphics[width=0.8\textwidth]{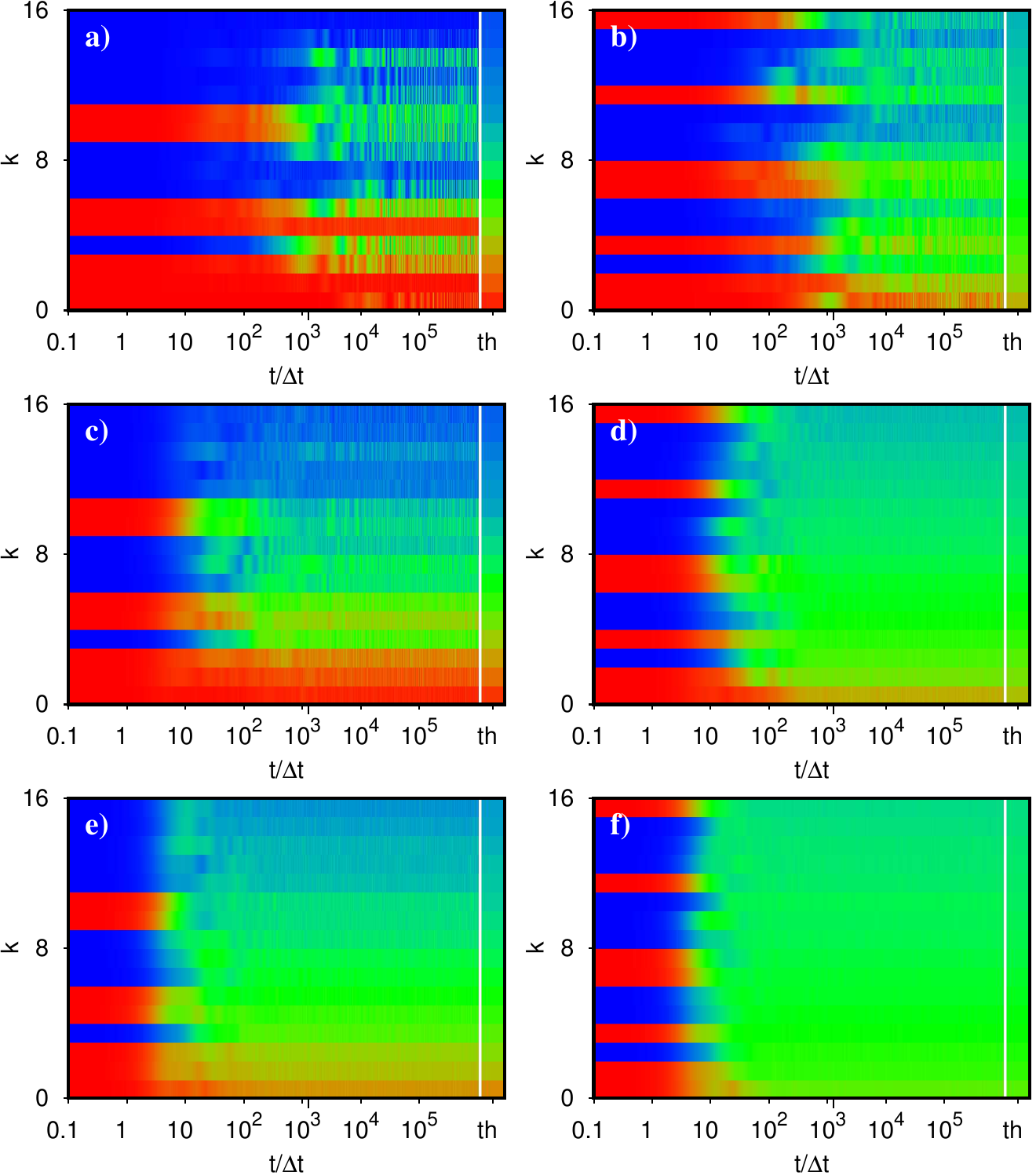}
\caption{\label{fig9} As in Figure~\ref{fig8} but 
for the initial states 
$|\psi(0)\!\!>=|\phi_3\!\!>=|0000011000110111\!\!>$ \aa, \cc, \ee\ and 
$|\psi(0)\!\!>=|\phi_4\!\!>=|1000100011001011\!\!>$ \bb, \dd, \ff\ (with 
same $A$ values as in Figure~\ref{fig8} for each row). 
These initial states can be obtained from the eigenstates of $H$ for 
$A=0.35$ at level numbers $m=123$ or $1354$ respectively 
by rounding the occupation numbers to 1 (or 0) if $n_k>0.5$ 
($n_k<0.5$) (see also top panels of Figure~\ref{fig2}). 
}
\end{figure}

The linear time scale used in Figure~\ref{fig7} is not very convenient 
since it cannot well capture a rapid decay/increase of $n_k$ at 
small times and its maximal time value is also significantly limited 
below the Heisenberg time. Therefore we use in Figures~\ref{fig8} and 
\ref{fig9} a logarithmic time scale with 
$0.1\,\Delta t\le t\le 10^6\,\Delta t\approx 10^2\, t_{\rm H}$. 
Note that in these figures the different $n_k$ values for each cell are not 
time averaged but represent the precise values for certain, 
exponentially increasing, discrete time values (see caption 
of Figure~\ref{fig8} for the precise values). Therefore in case of periodic 
oscillations of $n_k$ there will be, for larger time values, a quasi 
random selection of different time positions with respect to the period. 

In Figure~\ref{fig8} the time evolution for the initial states $|\phi_1\!\!>$ 
and $|\phi_2\!\!>$ is shown for the \Aberg values $A=1,\,3.5,\,10$. 
For $|\phi_1\!\!>$ at $A=1$ and $A=3.5$ the observations of Figure~\ref{fig7} 
are confirmed with the further information that the absence of thermalization 
in these cases is also valid for time scales larger than $2000\,\Delta t$ 
up to $10^6\,\Delta t$ and for $A=3.5$ the initial decay of $n_6$ and 
$n_{12}$ happens at $t\approx 10\,\Delta t$. For $|\phi_1\!\!>$ at $A=10$ 
the decay starts at $t\approx 3\,\Delta t$ and an approximate thermalization 
happens at $t>40\,\Delta t$. But here there is still some time periodic 
structure and it would be necessary to do some time average to have perfect 
thermalization. 
For $|\phi_2\!\!>$ at $A=1$ the decay of excited orbitals 12 and 14 starts 
at $t\approx 100\,\Delta t$ and saturates at $t\approx 1000\,\Delta t$ 
at which time also orbitals 6 and 7 are excited. After this there are 
very small excitations of orbitals 8, 9, 10 and maybe 13, 15. There is 
also some very modest decay of the Fermi sea orbitals 2, 4 and 5 at 
$t>1000\,\Delta t$. The final state at $t=10^6\,\Delta t$ is not 
thermalized even though some orbitals have $n_k$ values close to 
thermalization. For $|\phi_2\!\!>$ at $A=3.5$ the decay of excited 
orbitals 12 and 14 starts at $t\approx 10\,\Delta t$ and for 
$t>300\,\Delta t$ there is thermalization (but requiring some time average 
as for $|\phi_1\!\!>$ at $A=10$). Interestingly at intermediate times 
$10\,\Delta t<t<100\,\Delta t$ the high orbitals 13 and 16 are temporarily 
slightly excited and decay afterwards rather quickly to their thermalized 
values. For $|\phi_2\!\!>$ at $A=10$ the decay of excited 
orbitals 12 and 14 starts even at $t\approx 3\,\Delta t$ and 
thermalization seems to set in at $t>30\,\Delta t$.

Figure~\ref{fig9} is similar to Figure~\ref{fig8} but for the initial 
states $|\phi_3\!\!>$ and $|\phi_4\!\!>$ which have occupations numbers 
$n_k\in\{0,\,1\}$ obtained by rounding the $n_k$ values of the two eigenstates 
visible in the two top panels of Figure~\ref{fig2}. Here the initial decay 
of excited orbitals starts roughly at $t\approx 300\,\Delta t$ 
($t\approx (10-20)\,\Delta t$ or $t\approx (2-3)\,\Delta t$) for 
$A=1$ ($A=3.5$ or $A=10$ respectively). There is no thermalization 
for both states at $A=1$ (but some $n_k$ values are close to thermalized 
values), approximate thermalization for $A=3.5$ and $|\phi_3\!\!>$ 
and good thermalization for $A=3.5$ and $|\phi_4\!\!>$ as well as 
$A=10$ (both states). 

\begin{figure}[H]
\centering
\includegraphics[width=0.8\textwidth]{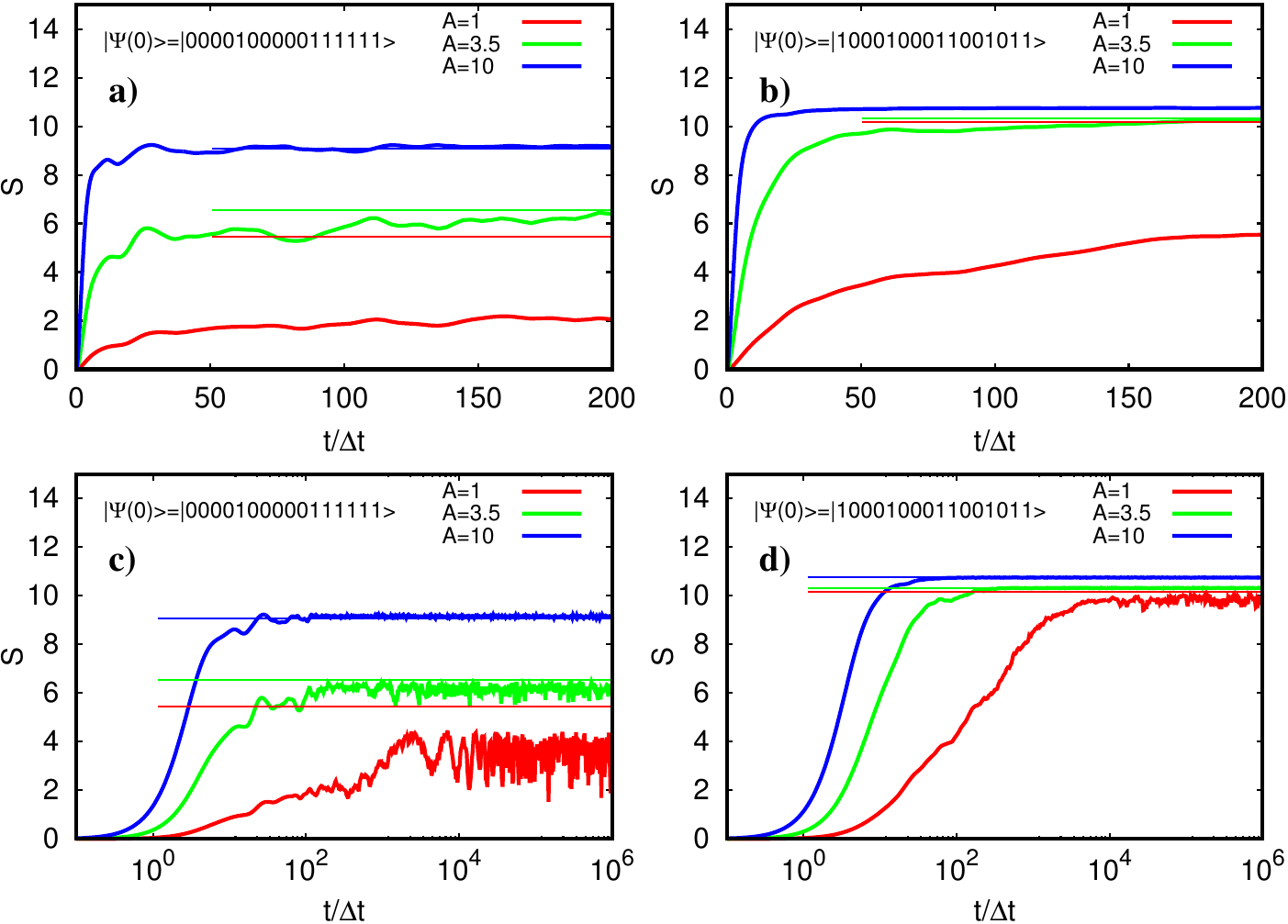}
\caption{\label{fig10} Time dependence of the entropy $S$, computed by 
(\ref{eq_entropy}), of the 
state $|\psi(t)\!\!>=\exp(-iHt)|\psi(0)\!\!>$ 
for the \Aberg parameter values $A=1$ (red lines), $A=3.5$ (green lines), 
$A=10$ (blue lines) and initial states 
$|\psi(0)\!\!>=|\phi_1\!\!>=|0000100000111111\!\!>$ \aa, \cc;
$|\psi(0)\!\!>=|\phi_4\!\!>=|1000100011001011\!\!>$ \bb, \dd;
thick colored lines show numerical data of $S(t)$ and thin horizontal 
colored lines show the thermalized entropy $S_{\rm th}(E_{\rm 1p})$ 
with $E_{\rm 1p}$ being determined from $|\psi(t)\!\!>$ at $t=10^6\,\Delta t$; 
panels \aa, \bb\ use 
a linear time axis: $0\le t\le 200\,\Delta t$; panels \cc, \dd\  
use a logarithmic time axis: $0.1\,\Delta t\le t\le 10^6\,\Delta t$; 
$\Delta t$ is the elementary quantum time step (see also 
Figure~\ref{fig6}). 
}
\end{figure}

Using the time dependent values $n_k(t)$ one can immediately determine 
the corresponding entropy $S(t)$ using (\ref{eq_entropy}). At $t=0$ we have 
obviously $S(0)=0$ since all four initially considered states we  have 
perfect occupation 
number values of either $n_k=0$ or $n_k=1$. Naturally one would expect 
that the 
entropy increases with a certain rate and saturates then at some maximal value 
which may correspond (or be lower) to the thermalized entropy 
$S_{\rm th}(E_{\rm 1p})$ 
(with $E_{\rm 1p}$ determined for the state $|\psi(t)\!\!>$ at large times) 
depending if there is presence (or absence) of thermalization according to 
the different cases visible in Figures~\ref{fig8} and \ref{fig9}. 
However, in absence of thermalization we see that there may also exist 
periodic oscillations with a finite amplitude at very long time scales. 

In Figure~\ref{fig10}, we show the time dependent entropy $S(t)$ for the two 
initial states $|\phi_1\!\!>$, $|\phi_4\!\!>$ and the three values $A=1$, 
$A=3.5$ and $A=10$ of the \Aberg parameter. For $A=10$ there is indeed a 
rather rapid saturation of the entropy of both states at a maximal value 
which is indeed close to the thermalized entropy $S_{\rm th}(E_{\rm 1p})$. 
We note that $E_{\rm 1p}$ is not conserved 
at strong interactions and that its initial value 
$E_{\rm 1p}\approx 30$ ($E_{\rm 1p}\approx 35$) at $t=0$ evolves 
to $E_{\rm 1p}\approx 33.5$ ($E_{\rm 1p}\approx 37$) at large 
times for $|\phi_1\!\!>$ ($|\phi_4\!\!>$) corresponding roughly 
to $S\approx S_{\rm th}(E_{\rm 1p})\approx 9.2$ ($10.8$) visible 
as thin blue horizontal lines in Figure~\ref{fig10}. 
For $A=3.5$ (or $A=1$) the thermalized entropy values, visible 
as thin green (red) lines, are lower as compared to the 
case $A=10$ due to different final $E_{\rm 1p}$ values. 
For $A=3.5$ and $|\phi_4\!\!>$ there is also saturation of $S$ to 
its thermalized value. 
For $A=3.5$ and $|\phi_1\!\!>$ there seems to be an approximate 
saturation at a quite low value $S\approx 6$ but with periodic fluctuations 
in the range $6\pm 0.3$. 
For $A=1$ and $|\phi_4\!\!>$ there is a quite late and approximate 
saturation with some fluctuations which is visible 
for $t>10^4\,\Delta t$ and with $S\approx 10\pm 0.2$. 
For $A=1$ and $|\phi_1\!\!>$ there is a late periodic regime for 
$t>10^3\,\Delta$ 
with a quite large amplitude $S\approx 3\pm 1$ and with 
$S_{\rm max}\approx 4$ significantly below the thermalized entropy 
$S_{\rm th}(E_{\rm 1p})\approx 5.5$. The panels using 
a normal (instead of logarithmic) time scale with $t\le 200\,\Delta t$ 
miss completely the long time limits for $A=1$ and might wrongly 
suggest that there is an early saturation at quite low values of $S$. 

The periodic (or quasi-periodic) time dependence of $n_k(t)$ or $S(t)$,
for the cases with 
lower values of $A$ and/or an initial state with lower energy, indicates that 
for such states only a small number ($2,\,3,\,\ldots$) of exact eigenstates 
of $H$ contribute mostly in the expansion of $|\psi(t)\!\!>$ in terms of these 
eigenstates. 

\begin{figure}[H]
\centering
\includegraphics[width=0.6\textwidth]{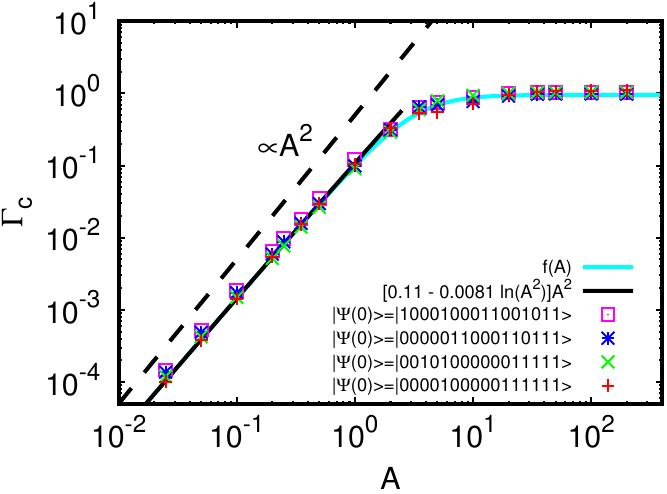}
\caption{\label{fig11} Dependence of the initial slope 
$\Gamma_c=\Delta t\,S'(t)$ (at small time values $t\sim \Delta t$) 
of the time dependent entropy on the \Aberg parameter 
$A$ in double logarithmic scale. 
$\Delta t$ is the elementary quantum time step (see also 
Figure~\ref{fig6}). The practical determination of $\Gamma_c$ 
is done using the fit 
$S(t)=S_\infty\,(1-\exp[-\gamma_1(t/\Delta t)-\gamma_2(t/\Delta t)^2])$
which provides $\Gamma_c=S_\infty\,\gamma_1$. The 
different data points correspond to the four different initial states 
used in Figures~\ref{fig8} and \ref{fig9}. The dashed line corresponds to the
power law behavior $\ \propto A^2$ and the light blue line corresponds 
to the fit $\Gamma_c=f(A)=(C_1-C_2\,\ln[g(A)])\,g(A)$ with 
$g(A)=A^2/(1+C_3 A^2)$ and fit values $C_1=0.107\pm 0.009$, 
$C_2=0.0081\pm 0.0023$, $C_3=0.092\pm 0.017$ for the initial state 
$|\psi(0)\!\!>=|\phi_1\!\!>=|0000100000111111\!\!>$ 
corresponding to the red plus symbols. Fit values for the other 
initial states can be found in Table~\ref{table2}.
The full black line corresponds to 
$f_0(A)=f(A)_{C_3=0}=[C_1-C_2\,\ln(A^2)]\,A^2$. The simpler fit 
$\Gamma_c=f_0(A)$ in the range $0.025\le A\le 1$ provides the 
values $C_1=0.107\pm 0.002$ and $C_2=0.0078\pm 0.0005$ which 
are identical (within error bars) 
to the values found by the more general fit $\Gamma_c=f(A)$ 
for the full range of $A$ values. 
}
\end{figure}

Figure~\ref{fig10} also shows that the initial increase of $S(t)$ is rather 
comparable between the two states for identical values of $A$ even though 
the long time limit might be very different. Furthermore a closer inspection 
of the data indicates that typically $S(t)$ is close to a quadratic behavior 
for $t\lesssim \Delta t$ but which immediately becomes linear 
for $t\gtrsim \Delta t$ similarly as the transition probabilities between 
states in the context of time dependent perturbation theory. 
To study the approximate slope in the linear regime we define 
\footnote{For practical reasons we decide to incorporate the quantum time 
step $\Delta t$ in the definition of $\Gamma_c$, i.e. $\Gamma_c$ 
is defined as the ratio of the initial slope $S'(t)$ over the 
global spectral bandwidth $\sim \sigma(A)\sim 1/\Delta t$.} 
the quantity $\Gamma_c=dS(t)/d(t/\Delta t)=\Delta t S'(t)$ 
for $t=\xi\Delta t$ where $\xi\gtrsim 1$ is a numerical constant 
of order one. To determine $\Gamma_c$ practically we perform 
first the fit $S(t)=\bar S_\infty\,(1-\exp[-\bar \gamma_1(t/\Delta t)])$ for 
$0\le t/\Delta t\le 100$ and use the exponential decay rate $\bar \gamma_1$ 
to perform a refined fit 
$S(t)=S_\infty\,(1-\exp[-\gamma_1(t/\Delta t)-\gamma_2(t/\Delta t)^2])$
for the interval $0\le t/\Delta t\le 5/\bar\gamma_1$. 
From this we determine $\Gamma_c=S_\infty\,\gamma_1$ which is rather 
close to $\bar S_\infty\,\bar\gamma_1$ for $A\le 2$ but not for larger values 
of $A$ where the decay time is reduced and not sufficiently large in 
comparison to the initial quadratic regime. Therefore the quadratic term in 
the exponential is indeed necessary to obtain a reasonable fit quality. 
This procedure corresponds to an effective average of the value of $\xi$ 
between $1$ and roughly $1/\gamma_1$ which is indeed useful to smear out some 
oscillations in the initial increase of $S(t)$ for smaller values 
$A\le 1$. 

Figure~\ref{fig11} shows the dependence of these values of $\Gamma_c$ 
on the parameter $A$ for our four initial states. At first sight 
on observes a behavior $\Gamma_c\propto A^2$ for $A\lesssim 2$ and 
a saturation for larger values of $A$. However, a more careful analysis 
shows that there are modest but clearly visible deviations with respect to 
the quadratic behavior in $A$ (power law fits $\Gamma_c\propto A^p$ 
for $A\le 2$ provide exponents close to $p\approx 1.75-1.85$) and it turns 
out that these deviations correspond to a logarithmic correction: 
$\Gamma_c=f(A)=(C_1-C_2\,\ln[g(A)])\,g(A)$ with $g(A)=A^2$ (for fits 
with $A\le 1$) or with $g(A)=A^2/(1+C_3\,A^2)$ (for fits with all $A$ values). 

To understand this behavior we write for sufficiently small times 
$n_k(t)\approx 1-\delta n_k(t)$ (if $n_k(0)=1$) or 
$n_k(t)\approx \delta n_k(t)$ (if $n_k(0)=0$) where 
$\delta n_k(t)$ is the small modification of $n_k(t)$. 
Time dependent perturbation theory 
suggests that $\delta n_k(t)\sim (t/\Delta t)^2$ for $t\lesssim \Delta t$ and 
$\delta n_k(t)\approx a_k\,A^2\,t/\Delta t$ for $t\gtrsim \Delta t$ such 
that still $\delta n_k(t)\ll 1$ with coefficients $a_k$ dependent on $k$ 
(and also on $M$, $L$) and satisfying 
a linear relation to ensure the conservation of particle number. 
Using (\ref{eq_entropy}) and neglecting corrections of order 
$\delta n_k^2$ we obtain: 
$S\approx -\sum_k\left(\delta n_k\ln \delta n_k-\delta n_k\right)$ 
and $\Gamma_c=\Delta t\,S'(t)\approx -\Delta t\sum_k \delta n_k'(t)
\ln \delta n_k(t)$ 
with $t=\xi\,\Delta t$. Since $\delta n_k'(t)\approx a_k A^2/\Delta t$ 
we find indeed the behavior~: 
\begin{equation}
\label{eq_Gamma_c_log}
\Gamma_c=[C_1-C_2\,\ln(A^2)]\,A^2\quad,\quad 
C_1=-\sum_k a_k\langle\ln(a_k\xi)\rangle\quad,\quad C_2=\sum_k a_k
\end{equation}
where $\langle\cdots\rangle$ indicates an average over some modest values 
of $\xi\gtrsim 1$. The precise values of $a_k$ may depend rather strongly 
on the orbital index $k$ and the initial state (see also 
Figures~\ref{fig8} and \ref{fig9}) but the coefficients $C_1$, $C_2$ depend 
only slightly on the initial state (see Table~\ref{table2}). 
Furthermore, by replacing $A^2\to g(A)=A^2/(1+C_3\,A^2)$ to allow 
for a saturation 
at large $A$ and with a further fit parameter $C_3$ it is possible to describe 
the numerical data by the more general fit $\Gamma_c=f(a)$ for the full 
range of $A$ values. 
\begin{table}[H]
\caption{\label{table2} Values of the fit parameters $C_1,\ldots,C_5$ 
for the initial states $|\phi_1\!\!>,\ldots,|\phi_4\!\!>$ used for the 
analytical fits of $\Gamma_c$ ($\Gamma_F$) in Figure~\ref{fig11} 
(Figure~\ref{fig13}). 
}

\centering
\include{table2pr}
\end{table}

\subsection{Survival probability and Fermi's golden rule}

\begin{figure}[H]
\centering
\includegraphics[width=0.8\textwidth]{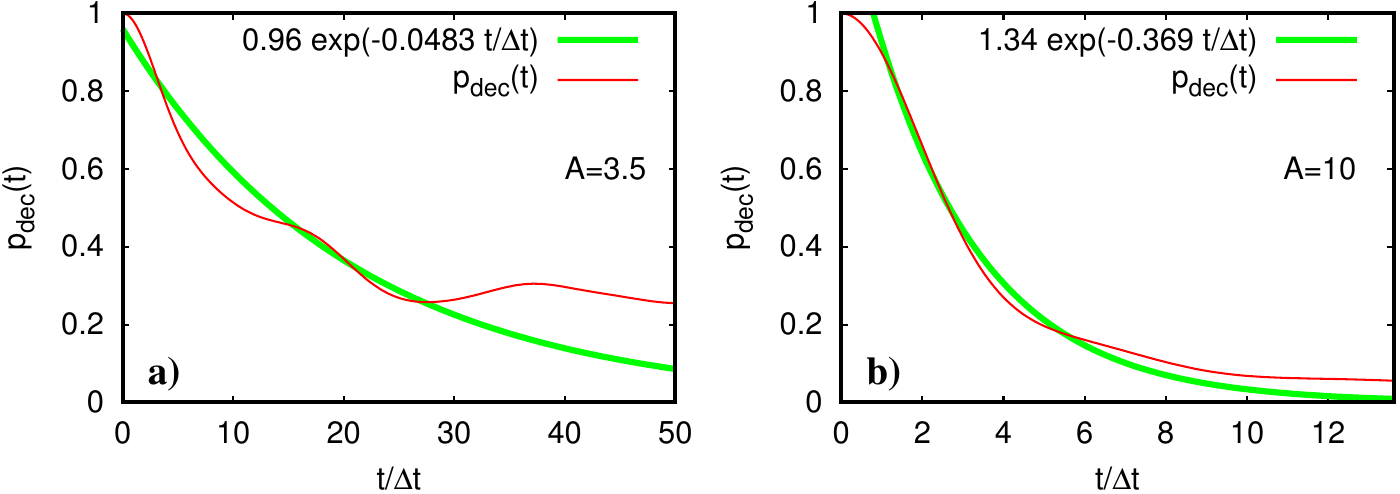}
\caption{\label{fig12} 
Decay function $p_{\rm dec}(t)=|<\!\!\psi(0)|\psi(t)\!\!>|^2$ obtained 
numerically from $|\psi(t)\!\!>$ with the initial state 
$|\psi(0)\!\!>=|0000100000111111\!\!>$ (thin red line) 
and the fit $p_{\rm dec}(t)=C\,\exp(-\Gamma_{\rm F}\, t/\Delta t)$ 
(thick green line)
for the two \Aberg values $A=3.5$ \aa\ and $A=10$ \bb. The fit values 
are $C=0.959\pm 0.011$, $\Gamma_{\rm F}=0.0483\pm 0.0015$ \aa; 
$C=1.339\pm 0.015$, $\Gamma_{\rm F}=0.369\pm 0.005$ \bb\ 
corresponding to the decay times 
$\Gamma_{\rm F}^{-1}=20.7$ \aa; $2.71$ \bb.
$\Delta t$ is the elementary quantum time step (see also 
Figure~\ref{fig6}). 
}
\end{figure}

The knowledge of the time dependent states $|\psi(t)\!\!>$ allows us 
also to compute the decay function 
$p_{\rm dec}(t)=|<\!\!\psi(0)|\psi(t)\!\!>|^2$ which represents the 
survival probability of the initial non-interacting eigenstate due 
to the influence of interactions. Again 
for the very short time window $t\lesssim \Delta t$ we expect a quadratic 
decay: $1-p_{\rm dec}(t)\approx \langle (H-E_{\rm mean})^2\rangle t^2
\approx {\rm const.}\,(t/\Delta t)^2$ with $\langle\cdots\rangle$ 
being the quantum expectation value with respect to $|\psi(0)\!\!>$ 
and a numerical constant $\lesssim 1$ since $1/\Delta t$ represents 
roughly the spectral width of $H$. For $t\gtrsim \Delta t$ but such that 
$1-p_{\rm dec}(t)\ll 1$ we have according to Fermi's golden rule:
$1-p_{\rm dec}(t)=\Gamma_F\,(t/\Delta t)$ where $\Gamma_F$ is the decay rate 
\footnote{\label{foot4} 
Again for practical reasons and similarly to $\Gamma_c$ we 
incorporate in the definition of $\Gamma_F$ the time scale $\Delta t$, i.e. 
$\Gamma_F=\Delta t\ \times$ usual decay rate found in the literature 
and meaning that $\Gamma_F$ is defined as the ratio of the usual 
decay rate over the global spectral bandwidth.}
of the state. 

To determine numerically $\Gamma_F$ we apply the fit: 
$p_{\rm dec}(t)=C\,\exp(-\Gamma_{\rm F}\, t/\Delta t)$ 
in two steps. First we use the interval $1\le t/\Delta t\le 50$ 
and if $5/\Gamma_F< 50$, corresponding to a rapid decay (which happens 
for larger values of $A$) we repeat the fit for the reduced interval 
$1\le t/\Delta t\le 5/\Gamma_F$. The choice of the Amplitude $C\neq 1$ 
and the condition $t\ge \Delta t$ for the fit range allow to take into 
account the effects due to the small initial window of quadratic decay.
In Figure~\ref{fig12} we show two examples 
for the initial state $|\phi_1\!\!>$ 
and the \Aberg values $A=3.5$ and $A=10$. In both cases the shown maximal 
time value $t_{\rm max}=50\,\Delta t$ (if $A=3.5$) or 
$t_{\rm max}\approx 13.5$ (if $A=10$) defines the maximal time value for 
the fit range. For $A=10$ the fit nicely captures the decay for 
$1\le t/\Delta t\le 6$ while for $A=3.5$ there are also some oscillations 
in the decay function for which the fit procedure is equivalent to some 
suitable average in the range $1\le t/\Delta t\le 30$. 
For very small values of $A$ the fit procedure works also correctly since 
it captures only the initial decay which is important if $p_{\rm dec}(t)$ 
does not decay completely at large times and which typically happens in the 
perturbative regime $A\lesssim 1$. 

\begin{figure}[H]
\centering
\includegraphics[width=0.6\textwidth]{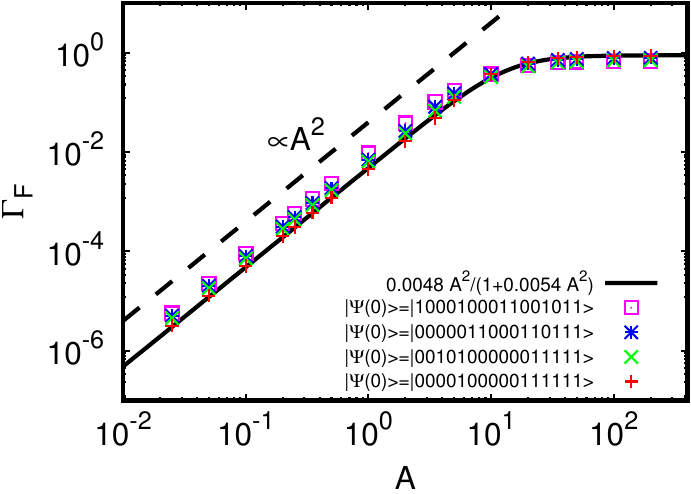}
\caption{\label{fig13} Dependence of the decay rate $\Gamma_F$ 
corresponding to Fermi's gold rule on the \Aberg parameter 
$A$ in double logarithmic scale. The practical determination of $\Gamma_F$ 
is done using the exponential fit function of Figure~\ref{fig12} for 
the numerically computed decay function $p_{\rm dec}(t)$. The 
different data points correspond to the four different initial states 
used in Figures~\ref{fig8} and \ref{fig9}. The dashed line corresponds to the
power law behavior $\ \propto A^2$ and the full black line corresponds 
to the fit $\Gamma_F=f(A)=C_4 A^2/(1+C_5 A^2)$ and fit values 
$C_4=0.0048\pm 0.0001$, $C_5=0.0054\pm 0.0003$ for the initial state 
$|\psi(0)\!\!>=|\phi_1\!\!>=|0000100000111111\!\!>$ 
corresponding to the red plus symbols. Fit values for the other 
initial states can be found in Table~\ref{table2}.
$\Delta t$ is the elementary quantum time step (see also 
Figure~\ref{fig6}). 
}
\end{figure}

Figure~\ref{fig13} shows the dependence of $\Gamma_F$ on $A$ for the usual 
four initial states together with the fit $\Gamma_F=f(A)=C_4 A^2/(1+C_5 A^2)$ 
for the data with initial state $|\phi_1\!\!>$. The values of the parameters 
$C_4$, $C_5$ for this and the other initial states are given in 
Table~\ref{table2}. Here the initial quadratic dependence 
$\Gamma_F\propto A^2$ is highly accurate (with no logarithmic correction). 
Similarly to $\Gamma_c$ there is only a slight dependence of the 
values of $\Gamma_F$ and the fit values on the choice of initial state. 

Theoretically, we expect according to Fermi's Golden rule that:
$\Gamma_F\approx (\Delta t)\,2\pi V_{\rm Fock}^2\, \rho_c(E)$ 
where $V_{\rm Fock}^2=\mbox{Tr}_{\rm Fock}(V^2)/(Kd)=\sigma_0^2\,\alpha A^2/K$
according to the discussion below (\ref{eq_sigma_int}) and $\rho_c(E)$ 
is the effective two-body density 
of states for states directly coupled by the interaction such that 
$\rho_c(E_{\rm mean})=1/\Delta_c$ (see discussion below (\ref{eq_Emean_var})). 
We note that $V_{\rm Fock}$ is the typical interaction matrix element in Fock 
space which is slightly larger than $V_{\rm mean}$ (see the theoretical 
discussion above for the computation of the coefficient 
$\alpha$ used in (\ref{eq_sigma_int}) and Appendix A of \cite{frahmtbrim}). 
The factor $\Delta t$ is due to our particular definition $^{\ref{foot4}}$ 
of decay rates. The expression of $\Gamma_F$ is actually also valid 
for larger values of $A$ provided we use the density of states 
$\rho_c(E)$ in the presence of interactions which provides an additional 
factor $1/\sqrt{1+\alpha A^2}$ according to (\ref{eq_sigma_int}). 
Therefore, at the band center we have: 
$2\pi\,\Delta t\,\rho_c=K/[\sigma_0(L)\,\sigma_0(L=2) (1+\alpha A^2)]$ 
which gives together with (\ref{eq_sigam2p}): 
\begin{equation}
\label{eq_GammaF}
\Gamma_F=
\frac{\sigma_0(L)}{\sigma_0(L=2)}\left(\frac{\alpha A^2}{1+\alpha\,A^2}\right)=
\sqrt{\frac{L(M-L)}{2(M-2)}}\left(\frac{\alpha A^2}{1+\alpha\,A^2}\right).
\end{equation}
For $M=16$ and $L=7$ the square root factor is 1.5 and we have to 
compare $1.5\alpha\approx 0.0132$ with the values of $C_4$ in 
Table \ref{table1} which are somewhat smaller, probably due to a reduction 
factor for the energy dependent density of states since the energies of the 
initial states have a certain distance to the band center. 
Furthermore, according to (\ref{eq_GammaF}), we have to compare $C_5$ with 
$\alpha\approx 0.00877$ which is not perfect but gives the correct 
order of magnitude. For both parameters the numerical matching is quite 
satisfactory taking into account the very simple argument using the same 
typical value of the interaction matrix elements for all cases of 
initial states. 

Finally, we mention that for the three \Aberg parameter values 
$A=1$, $A=3.5$, $A=10$ used in Figures~\ref{fig8} and \ref{fig9} we have 
typical decay times in units of $\Delta t$ being 
$1/\Gamma_F\approx 300$, $30$, $3$ respectively (with some modest 
fluctuations depending on initial states). These values match quite well 
the observed time values at which the initially occupied orbitals start 
to decay (see above discussion of Figures~\ref{fig8} and \ref{fig9}). 

\subsection{Time evolution of density matrix and spatial density}

\begin{figure}[H]
\centering
\includegraphics[width=0.8\textwidth]{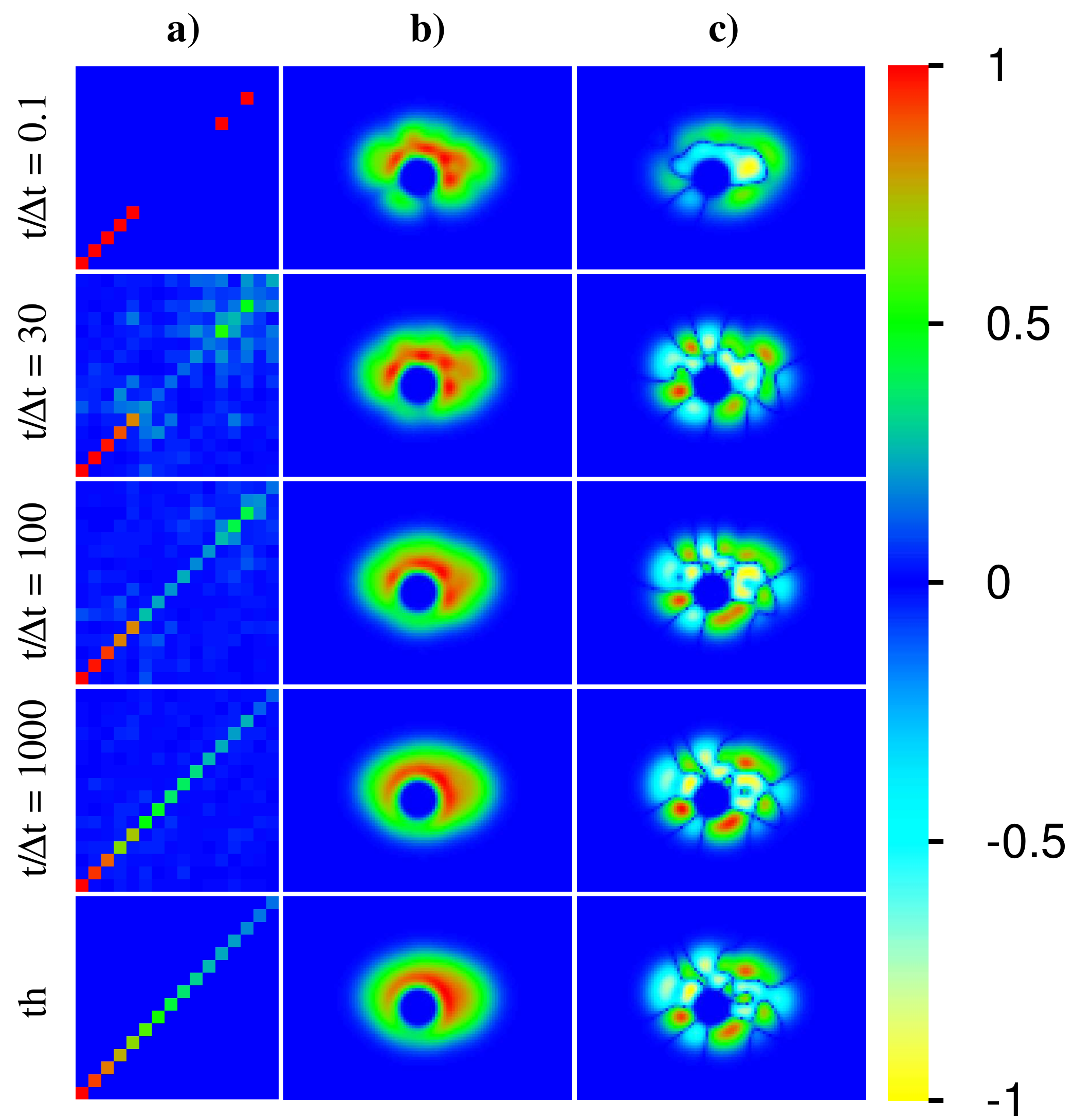}
\caption{\label{fig14} 
Time dependent density matrix $|n_{kl}(t)|$ \aa, 
spatial density $\rho(x,y,t)$ \bb, 
spatial density difference with respect to the initial condition 
$\Delta \rho(x,y,t)=\rho(x,y,t)-\rho(x,y,0)$ \cc\ all computed 
from $|\psi(t)\!\!>$ for the initial state 
$|\psi(0)\!\!>=|\phi_2\!\!>=|0010100000011111\!\!>$ 
and the \Aberg parameter $A=3.5$. 
Panels in column \aa\ corresponds to $(k,l)$ plane with 
$k,l\in\{1,\dots,\,16\}$ 
being orbital index numbers. Panels in columns \bb\ and \cc\  correspond to 
the same rectangular domain in $(x,y)$ plane as in Figure~\ref{fig1}. 
The five rows of panels correspond to the time values 
$t/\Delta t=0.1,\,30,\,100,\,1000$ and the 
thermalized case (label ``th'') where the density matrix is diagonal 
with entries being the thermalized occupations numbers 
$n_{kk}=n(\eps_k)$ at energy $E=32.9$ (typical total one-particle energy 
of $|\psi(t)\!\!>$ for $t/\Delta t \ge 1000$). 
The numerical values of the color bar
represent values of $\ |n_{kl}|$ \aa, $\ (\rho/\rho_{\rm max})^{1/2}$ 
\bb, $\ \sgn(\Delta\rho)(|\Delta\rho|/\Delta\rho_{\rm max})^{1/2}$ \cc\ where 
$\rho_{\rm max}$ or $\Delta\rho_{\rm max}$ are maximal values of 
$\rho$ or $|\Delta\rho|$ respectively. 
}
\end{figure}

We now turn to the effects of the many-body time evolution in position 
space (see for example Figure~\ref{fig1}). For this we compute the 
spatial density 
\begin{equation}
\label{eq_MBdensity_pos}
\rho(x,y,t)=<\!\!\psi(t)|\Psi^\dagger (x,y)\,\Psi (x,y) |\psi(t)\!\!>
\quad,\quad \Psi^{(\dagger)}(x,y)=
\sum_k\,\varphi^{(*)}_k(x,y)\,c^{(\dagger)}_k
\end{equation}
where $\varphi_k(x,y)$ is the one-particle eigenstate of orbital $k$, with 
some examples shown in Figure~\ref{fig1}. Here $\Psi^{(\dagger)}(x,y)$ 
denote the usual fermion field operators (in case of continuous $x$, $y$ 
variables) or standard fermion operators for discrete position 
basis states (when using a discrete grid for $x$ and $y$ positions 
as we did for the numerical solution of the non-interacting 
Sinai oscillator model in Section~\ref{sec2}). The sum over orbital index $k$ 
in (\ref{eq_MBdensity_pos}) requires in principle a sum over a 
{\em full complete} basis set of orbitals with infinite number 
(case of continuous $x$, $y$ values) or a very large number 
(case of discrete $x$-$y$ grid) significantly larger than the very modest 
number of orbitals $M$ we used for the numerical 
solution of the many-body Sinai oscillator.

However, we can simply state that in our model, by construction, all 
orbitals with $k>M$ are never occupied such that in the expectation 
value for $\rho(x,y,t)$ only the values $k\le M$ are necessary. 
Taking this into account together with the fact that the one-particle 
eigenstates are real valued, we obtain the more explicit expression:
\begin{equation}
\label{eq_dens_explicit}
\rho(x,y;t)=\sum_{k,l=1}^M \varphi_k(x,y)\,\varphi_l(x,y)\,
n_{kl}(t)\quad,\quad 
n_{kl}(t)=<\!\!\psi(t)| c^{\dagger}_k\,c^{\pdag}_l|\psi(t)\!\!>
\end{equation}
where $n_{kl}(t)$ is the density matrix in orbital representation 
generalizing the occupation numbers $n_k(t)$ which are its diagonal elements. 
Due to the complex phases of $|\psi(t)\!\!>$ (when expanded in the 
usual basis of non-interacting many-body states) the density matrix is 
complex valued but hermitian: $n_{kl}^*(t)=n_{lk}(t)$. Therefore its 
anti-symmetric imaginary part does not contribute in $\rho(x,y;t)$. 
We have numerically evaluated (\ref{eq_dens_explicit}) and we present 
in Figure~\ref{fig14} color plots of the density matrix and the 
spatial density $\rho(x,y;t)$ for $A=3.5$, 
the initial state $|\psi(0)\!\!>=|\phi_2\!\!>$ 
and four time values $t/\Delta t=0.1,\,30,\,100,\,1000$. Since the 
density $\rho(x,y;t)$ does not provide a lot of spatial structure 
we also show in Figure~\ref{fig14} the density difference with respect to 
the initial condition $\Delta \rho(x,y;t)=\rho(x,y;t)-\rho(x,y;0)$ which 
reveals more of its structure 
(figures and videos for the time evolution of this and other cases are 
available for download at the web page \cite{quantware}). 

At the time $t/\Delta t=0.1$ density matrix and spatial density are 
essentially identical to the initial condition at $t=0$. For 
$\Delta\rho$ we see a non-trivial structure since there is a small 
difference with the initial condition and the color plot simply 
amplifies small maximal amplitudes to maximal color values (red/yellow for 
strongest positive/negative values even if the latter are small in an 
absolute scale). The density matrix is diagonal 
and its diagonal values are either $1$ (for initially occupied orbitals) 
or $0$ (for initially empty orbitals) and the spatial density 
simply gives the sum of densities due to the occupied eigenstates. 

At $t/\Delta t=30$ we see a non-trivial structure in the density matrix 
with a lot of non-vanishing values in certain off-diagonal elements. 
Furthermore the orbitals 13 and 16 are also slightly excited (see 
also discussion of Figure~\ref{fig8}) and there is a significant 
change of the spatial density. 

Later at $t/\Delta t=100$ the number/values of off-diagonal elements 
in the density matrix is somewhat reduced but they are still visible. 
Especially between orbitals 12 and 13 as well as 14 and 15 there is a 
rather strong coupling. 
Orbital 13 is now stronger excited than the initially excited orbital 12. 
Also orbital 14 and 15 are quite strong. The spatial density has become 
smoother and the structure of $\Delta \rho$ is roughly close to the 
case at $t/\Delta t=30$ but with some significant differences. 

Finally at $t/\Delta t=1000$ the density matrix seems be diagonal with 
values close to the thermalized values. There is a further increase of the 
density smoothness and $\Delta \rho$ has a similar but 
different structure as for $t/\Delta t=100$ or $t/\Delta t=30$.

Apparently at intermediate times $20\le t/\Delta t\le 100$ there are some 
quantum correlations between certain orbitals, visible as off-diagonal 
elements in the density matrix which disappear for later times. This kind 
of decoherence is similar to the exponential decay observed in 
\cite{frahmtbrim} for the off-diagonal element of the $2\times 2$ density 
matrix for a qubit coupled to a chaotic quantum dot or the SYK black hole. 
However, to study this kind of decoherence more carefully in the context here 
it would be necessary to use as initial state a non-trivial linear combination 
of two non-interacting eigenstates and not to rely on the creation of 
modest off-diagonal elements for intermediate time scales as we see here. 

The spatial density is globally rather smooth and typically quite well 
given by the ``classical'' relation 
$\rho(x,y;t)\approx \sum_k \varphi_k^2(x,y)\,n_k(t)$ in terms of the 
time dependent occupation numbers. Only for intermediate time 
scales with more visible quantum coherence (more off-diagonal 
elements $n_{kl}(t)\neq 0$), this relation is less accurate. 
However, at $A=3.5$ the density still exhibits small but regular fluctuations 
in its detail structure as can be seen in the structure of $\Delta\rho$ 
for later time scales. A closer inspection of the data (for time values 
not shown in Figure~\ref{fig14}) also shows that even at long time scales 
there are significant fluctuations of $\rho$ when $t$ is slightly changed 
by a few multiples of $\Delta t$. 

In Figure~\ref{fig14} we also show for comparison the theoretical 
thermalized quantities where in (\ref{eq_dens_explicit}) 
the density matrix is replaced by a diagonal matrix with entries being the 
thermalized occupations numbers $n_{kk}=n(\eps_k)$ at energy $E=32.9$ 
which is the typical total one-particle average energy of $|\psi(t)\!\!>$ 
for the long time limit $t/\Delta t \ge 1000$ showing that at 
$t/\Delta t = 1000$ 
the state is very close to thermalization but still with small significant 
differences (see also discussion of Figure~\ref{fig7} for this case).

\begin{figure}[H]
\centering
\includegraphics[width=0.8\textwidth]{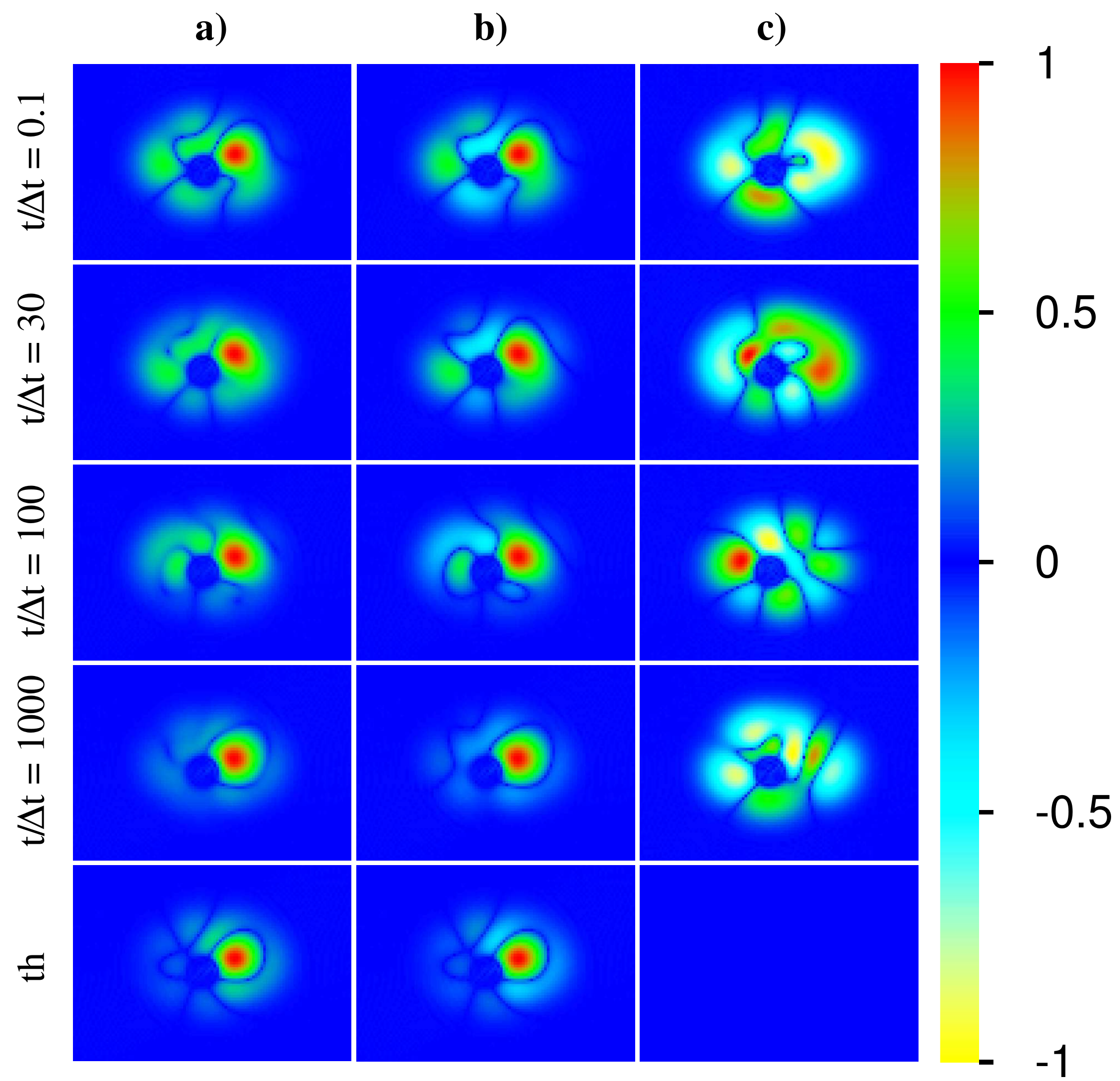}
\caption{\label{fig15} 
Time dependent spatial density correlator $\rho_{\rm corr}(x,y;x_0,y_0;t)$ 
shown in the same rectangular domain in $(x,y)$ plane as in Figure~\ref{fig1}. 
The columns correspond to absolute value \aa, real part \bb, 
imaginary part \cc. The initial point is given by 
$(x_0,y_0)=(1.22,0.15)$ which is very close to the maximal position (center 
of the red area) of the one-particle ground state $\varphi_1(x,y)$ visible 
in panel \aa\ of Figure~\ref{fig1}. Initial state, \Aberg parameter 
and meaning of row labels are as in Figure~\ref{fig14}. 
The numerical values of the color bar
represent values of $\ \sgn(u)|u/u_{\rm max}|^{1/2}\ $ where 
$u$ is absolute value \aa, real part \bb, 
imaginary part \cc, of $\rho_{\rm corr}$ and $u_{\rm max}$ is the 
maximal value of $|u|$. The data for thermalized case and imaginary part is 
completely zero (blue panel in bottom right corner) 
since the spatial density correlator is for the thermalized case purely real. 
}
\end{figure}

We may also generalize the spatial density (\ref{eq_MBdensity_pos}) 
to a spatial density correlator which we define as:
\begin{equation}
\label{eq_dens_corr}
\rho_{\rm corr}(x,y;x_0,y_0;t)=
<\!\!\psi(t)|\Psi^\dagger (x,y)\,\Psi (x_0,y_0) |\psi(t)\!\!>=
\sum_{k,l=1}^M \varphi_k(x,y)\,\varphi_l(x_0,y_0)\,
n_{kl}(t)
\end{equation}
depending on initial $(x_0,y_0)$ and final position $(x,y)$. 
As an illustration we choose the fixed value $(x_0,y_0)=(1.22,0.15)$ 
which is very close to the maximal position (center 
of the red area) of the one-particle ground state $\varphi_1(x,y)$ visible 
in panel \aa\ of Figure~\ref{fig1}. The spatial density correlator 
is potentially 
complex with a non-vanishing imaginary part in case of non-vanishing 
off-diagonal matrix elements of $n_{kl}(t)\neq 0$ for $k\ne l$. 
In Figure~\ref{fig15} we present density plots 
of absolute value, real and imaginary part of 
$\rho_{\rm corr}(x,y;x_0,y_0;t)$ in $(x,y)$ plane and with the given value 
$(x_0,y_0)=(1.22,0.15)$ for the same parameters 
of Figure~\ref{fig14} (concerning initial state, \Aberg parameter, time 
values and also thermalized case). However, for the thermalized case 
the density matrix is diagonal by construction and the imaginary 
part of $\rho_{\rm corr,th}(x,y;x_0,y_0)$ vanishes (giving a blue panel due 
to zero values). 

There are significant time dependent fluctuations of 
$\rho_{\rm corr}(x,y;x_0,y_0;t)$ for all time scales with real part and 
absolute value being dominated by rather strong maximal values for 
positions close to the initial position. However, the imaginary part 
(which vanishes at $t=0$ and is typically smaller than the values 
in maximum domain of real part) 
shows a more interesting structure since the color plot 
amplifies small amplitudes (in absolute scale). Apart from this the 
absolute and real part values for positions outside the maximum domain 
(far away from the initial position) seem to decay for long timescales which is 
also confirmed by the thermalized case. Even though the case for 
$t/\Delta t=1000$ seems to be rather close to the thermalized case 
(for absolute value and real part) there are still differences which 
are more significant here as in Figure~\ref{fig14}. 

In global the obtained results show that the dynamical thermalization 
well takes place leading to the usual Fermi-Dirac thermal distribution
when the \Aberg criterion is satisfied and interactions are 
sufficiently strong to drive the system into the thermal state. 

\section{IV Estimates for cold atom experiments}

We discuss here  typical parameters for cold  atoms in a trap following \cite{ketterlefermi}.
Thus for sodium atoms we have 
$\omega  \approx \omega_{x,y,z} \approx 2\pi 10 Hz$ with 
$a_0= \sqrt{\hbar/(m \omega)} = 6.5 \mu m$
and oscillator level spacing $E_u = \hbar \omega  \approx 0.5 n K$ (nanoKelvin).  
The typical scattering length is $a_s \approx 3 nm$ being small compared to $a_0$. 
The atomic density  is $\rho_0 =1/(a_0)^3 \approx 4 \times 10^9 cm^{-3}$.
Since $a_s \ll a_0$ the two-body interaction is of  $\delta$-function type
with $v({{\bf r_1-r_2}}) = (4\pi \hbar^2 a_s/m) \delta({\bf r_1-r_2})$ \cite{wilkens,ketterlefermi}. 
In our numerical simulations $\delta({\bf r})$ is replaced by a function 
$H(r_c-|{\bf r}|)/(C\,2^d r_c^d)$ with a small $r_c$,  volume $C$ of the unit 
sphere in $d$ dimensions and with the Heaviside function $H(x)=1$ (or $0$) 
for $x \ge 0$ and $H=0$ for  $x<0$. Hence, our 
parameter $U$ introduced in Section \ref{sec3} corresponds to 
$U=(C\,2^d r_c^d) (4\pi \hbar^2 a_s/m)$. 

Below we present the estimates for the dynamical thermalization border
for excitations of fermionic atoms in a vicinity of their Fermi energy in 3D Sinai oscillator
following the lines of Eq.(\ref{eq_g23}).
In such a case the two-body interaction energy scale 
between atoms is $U_s = 4\pi \hbar^2 \rho_0 a_s/m = 4\pi (a_s/a_0) \hbar \omega$ \cite{wilkens,ketterlefermi}
so that $U_s/\hbar \omega \sim 6 \times 10^{-3}$. 
Comparing to sodium the mass of Li atoms is
approximately 3 times smaller so that for the same $\omega$ we have $a_0 \approx 10 \mu m$
and  $U_s/\hbar \omega \approx 4 \times 10^{-3}$. We think that the scattering length
can be significantly increased via the Feshbach resonance allowing to reach
effective interaction values $U_s/\hbar \omega \sim 1$ being similar to the value $A \sim 3$
used in our numerical studies with the onset of dynamical thermalization.  

Usually a 3D trap with fermionic atoms can capture about $N_a \sim 10^5$ atoms
with $\omega  \approx \omega_{x} \sim \omega_{y} \sim \omega_{z} \sim 2\pi 10 Hz$. 
Following the result (\ref{eq_g23}) it is interesting to determine 
the DTC border dependence on   $N_a \gg 1$ for a Sinai oscillator 
with $r_d \sim 1 \mu m \sim  a_0/5$.
We assume that similar to 2D case 
the scattering on an elastic ball in the trap center  leads to 
quantum chaos and chaotic eigenstates with $\ell \leq N_a$ components 
(e.g. in the basis of oscillator eigenfunctions).
The Fermi energy of the trap is then 
$E_F = \hbar (N_a \omega_x \omega_y \omega_z)^{1/3} \approx \hbar \omega {N_a}^{1/3}$ 
\cite{roati1,roati2}.
Assuming that all these components have random amplitudes of a typical size 
$1/\sqrt{\ell}$ we then obtain an estimate for a typical matrix element
of two-body interaction between one-particle eigenstates 
\begin{equation}
\label{eq_u2trap}
U_2 \approx \alpha_s \hbar \omega/ \ell^{3/2} \; , \;\; \alpha_s = 4\pi (a_s/ a_0) \; , 
\;\; a_0 = \sqrt{\hbar/m \omega} \; . 
\end{equation}
The derivation of this estimate is very similar to the case of two interacting particles 
in a disordered potential with localized eigenstates \cite{tip}. At the same time
in a vicinity of the Fermi energy $E_F$ we have the one-particle level spacing
$\Delta_1 = d E_F/d N_a \approx \hbar \omega /(3 {N_a}^{2/3})$. 
Hence the effective conductance appearing
in (\ref{eq_g23}) is $g =\Delta_1 /U_2 \approx   \ell^{3/2}/ (3 \alpha_s {N_a}^{2/3} )$. Thus from 
(\ref{eq_g23}) we obtain the dynamical thermalization border 
for excitation energy $\delta E$
in a 3D Sinai-oscillator trap with $N_a$ fermionic atoms:
\begin{equation}
\label{eq_dtc3d}
 \delta E > \delta E_{ch} \approx \Delta_1 g^{2/3} \approx  
2  \ell \Delta_1 /({\alpha_s}^{2/3} {N_a}^{4/9}) 
\sim  {N_a}^{5/9} \Delta_1/{\alpha_s}^{2/3} \sim \hbar \omega /({\alpha_s}^{2/3} {N_a}^{1/9}) 
\sim E_F/({\alpha_s}^{2/3} {N_a}^{4/9}) \; .
\end{equation}
It is assumed that $\delta E \ll E_F$.
Here the last three relations are written in an assumption that $\ell \sim N_a$.
Thus at large $N_a$ values and not too small $\alpha_s$ the critical 
energy border $\delta E_{ch}$ for dynamical thermalization is rather small
compared to $E_F$.  However, still $\delta E_{ch} \gg \Delta_1$.
Here we used the maximal value for the number of components $\ell \sim N_a$.
It is possible that in a reality $\ell$ can be significantly smaller than $N_a$.
However, the determination of the dependence $\ell(N_a)$ 
requires separate studies taking into account the properties
of chaotic eigenstates and their spreading over the energy surface.
These spreading can have rather nontrivial properties (see e.g. \cite{rough2}). 
This is confirmed by the results presented in Appendix  
for the 2D case of Sinai oscillator showing 
the numerically obtained dependence of two-body matrix element on energy
for transitions in a vicinity of Fermi energy $E_F$ (see Fig.~\ref{figA1} there).

\section{V Discussion}

In this work we demonstrated the existence of interaction 
induced dynamical thermalization of fermionic atoms in a Sinai-oscillator trap
if the interaction strength between atoms exceeds 
a critical border determined by the \Aberg criterion 
\cite{aberg1,aberg2,jacquod}.
This thermalization takes place in a completely isolated system
in absence of any contact with an external thermostat.
In the context of the Loschmidt-Boltzmann dispute \cite{loschmidt,boltzmann}
we should say that formally this thermalization is
reversible in time since the Schrodinger equation of the system
has symmetry $t \rightarrow -t$. The classically chaotic dynamics 
of atoms in the Sinai-oscillator trap breaks in practice this reversibility
due to exponential growth of errors induced by chaos.
In the regime of quantum chaos there is no 
exponential growth of errors due to the fact that the Ehrenfest time scale 
of chaos is logarithmically short 
\cite{chirikov1981,chirikov1988,dls1981,stmapscholar}.
An example for the stability of time reversibility
is given in \cite{dls1983,stmapscholar}. In fact the experimental reversal 
of atom waves in the regime of quantum chaos has been even 
observed with cold atoms in \cite{hoogerland}.
In view of this and the fact that the 
spectrum of atoms in the Sinai-oscillator trap is discrete 
we can say that dynamical thermalization
will have obligatory revivals in time
returning from the thermalized state 
(e.g.  bottom panels in Figure \ref{fig8})
to the initial state (top panels in Figure \ref{fig8}).
This is the direct
consequence of the Poincare recurrence theorem
\cite{poincare}. However, the time for such a recurrence 
grows exponentially with the number of components 
contributing to the initial state 
(which is also exponentially large in the 
regime of dynamical thermalization) 
and thus during such a long time scale
external perturbations (coming from outside
of our isolated system, e.g. not perfect isolation)
will break in practice this time reversibility.

We hope that our results will initiate experimental
studies of dynamical thermalization with 
cold fermionic atoms in systems such as the Sinai-oscillator trap.

\section*{Acknowledgments}

We are thankful to Shmuel Fishman for deep discussions of quantum chaos problems 
and related scientific topics during many years.

{\it NOTE: This work represents the contribution to the Special Issue
in memory of Shmuel Fishman to appear at MDPI Condenced Matter journal.}

This work was supported in 
part by the Programme Investissements
d'Avenir ANR-11-IDEX-0002-02, 
reference ANR-10-LABX-0037-NEXT (project THETRACOM).
This work was granted access to the HPC GPU resources of 
CALMIP (Toulouse) under the allocation 2018-P0110.

\section*{A1. Two-body matrix element near the Fermi energy}
\label{sisec1}

In this Appendix we present numerical results
for the dependence of the quantity 
$V_{\rm mean}(\eps)=\sqrt{\langle V_{ij,kl}^2\rangle}$
as a function of $\eps$ where the average is done {\em only 
for orbitals with energies $\eps_n$ close to $\eps$} 
(for $n\in\{i,j,k,l\}$), i.e.:
$|\eps_n-\eps|\le \Delta\eps$ with $\Delta\eps=2$. We note that this 
is different from the quantity $V_{\rm mean}$ used in Section \ref{sec3} 
where the average was done over all orbitals (up to a maximal number being 
$M$). The reason for the special average with orbital energies close 
to $\eps$ (which will be identified with the Fermi energy $E_F$) 
is that these transitions are dominant in presence of the 
Pauli blockade near the Fermi level.

We remind that according to the discussion of Sections \ref{sec2} and
\ref{sec3} the matrix elements $V_{ij,kl}$ were computed for an interaction 
potential of amplitude $U$ for $|{\bf r_1-r_2}|<r_c$ 
(with the radius $r_c =0.2 r_d =0.2$) and being zero for 
$|{\bf r_1-r_2}|\ge r_c$. Furthermore, they 
have been anti-symmetrized and a 
diagonal shift $V_{ij,ij}\to V_{ij,ij}-(1/M_2)\sum_{k<l} V_{kl,kl}$ was 
applied to ensure that the interaction matrix has a vanishing trace. 

Due to this shift and the precise average procedure there is a slight 
(purely theoretical) dependence on the maximal orbital number $M$ 
for this average (there is a cut-off effect for $\eps$ close to the 
maximal orbital energy $\eps_M$). Due to this we considered two values 
of $M=30$ and $M=60$.

\begin{figure}
\includegraphics[width=0.50\columnwidth]{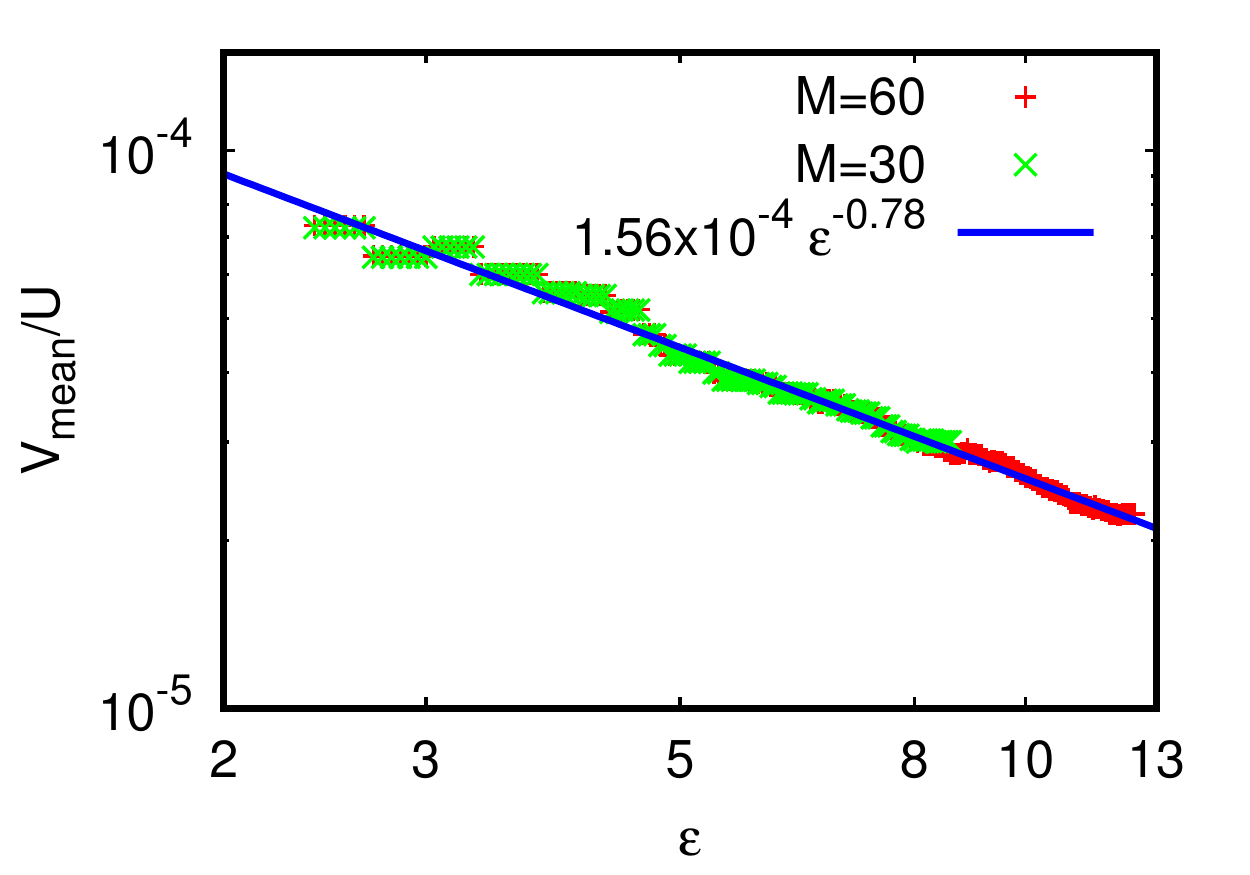}
\caption{Dependence of the average two-body matrix element
$V_{\rm mean}$ rescaled by the amplitude of interaction strength $U$
on one-particle energy $\eps$ for two-body interaction
transitions in a vicinity of Fermi energy $E_F=\eps$;
green symbols are for number of one-particle orbitals $M=30$,
red symbols are for $M=60$; the blue line shows the fit 
$V_{\rm mean}/U = a/\eps^b$ with
$a= 0.000156 \pm 2 \times 10^{-6}$;
$b= 0.781 \pm 0.005$.
}
\label{figA1}
\end{figure}

The numerically obtained dependence is shown in Fig.~\ref{figA1} 
and is well described by the fit $V_{\rm mean}/U = a/ \eps^b$ with  
$a= 1.56 \times 10^{-4}$ and $ b =0.78$. 
The small value of $a$ is due to antisymmetry of two-particle fermionic states
and due to a small value of $r_c=0.2 r_d$ which leads to a decrease
of the effective interaction strength being proportional to ${r_c}^2$.

We note that the Fermi energy is determined by the number of fermionic 
atoms $N_a$ inside the 2D Sinai oscillators
with $\eps=E_F \approx \omega {N_a}^{1/2}$ assuming
$\omega = \omega_x \approx \omega_y$.
Therefore we have  $\eps \propto M^{1/2} \sim {N_a}^{1/2}$,
$\Delta_1 \sim \hbar \omega /{N_a}^{1/2}$ and 
$V_{\rm mean} \sim \alpha_s \hbar \omega  / \ell^{3/2} \sim  \alpha_s \hbar \omega /{N_a}^{b/2}$ 
(see (\ref{eq_u2trap})).
Hence, the obtained exponent $b \approx 0.78$ corresponds to the number
of one-particle components 
$\ell \sim {N_a}^{b/3} \sim {N_a}^{0.25} \sim {n_x}^{0.5}$.  
At the moment we do not have a clear explanation for this numerical dependence.
This dependence corresponds to 
$g = \Delta_1 /V_{\rm mean} \sim \ell^{3/2}/ (\alpha_s {N_a}^{1/2}) \sim 1/(\alpha_s {N_a}^{1/8})$.
For such a dependence we obtain that the DTC border in 2D takes place for an excitation energy
$\delta E > \delta E_{ch} \sim g^{2/3} \Delta_1 \sim \hbar \omega /({\alpha_s}^{2/3} {N_a}^{7/12})$.
Thus the thermalization can take place at rather low energy excitations 
above the Fermi energy with $\Delta_1 < \delta E \ll E_F$.

\section{References}






\end{document}

%% file: table1pr.tex
\begin{tabular}{ccccccccccc}
\toprule
$A$ & $m$ & $S$ & $S_{th}(E_{\rm 1p})$ &  $E_{\rm 1p}$ & $\mu(E_{\rm 1p})$ & $\beta(E_{\rm 1p})$ &  $S_{\rm th}(E_{\rm ex})$ & $E_{\rm ex}$ & $\mu(E_{\rm ex})$ &  $\beta(E_{\rm ex})$ \\
0.35 & 122 & 0.95 & 7.91 & 32.15 & 5.31 & 1.05 & 7.89 & 32.13 & 5.31 & 1.05 \\ 
0.35 & 1353 & 4.91 & 10.16 & 35.29 & 4.98 & 0.45 & 10.16 & 35.30 & 4.98 & 0.45 \\ 
3.5 & 122 & 6.99 & 8.28 & 32.52 & 5.28 & 0.95 & 7.54 & 31.81 & 5.34 & 1.15 \\ 
3.5 & 1353 & 10.16 & 10.23 & 35.45 & 4.95 & 0.43 & 10.10 & 35.15 & 5.00 & 0.47 \\ 
10 & 122 & 8.91 & 8.98 & 33.33 & 5.22 & 0.77 & 4.96 & 30.10 & 5.46 & 2.02 \\ 
10 & 1353 & 10.52 & 10.54 & 36.28 & 4.75 & 0.32 & 9.53 & 34.12 & 5.14 & 0.63 \\ 
\end{tabular}

%% file: table2pr.tex
\begin{tabular}{cccccc}
\toprule
initial state & $C_1$ & $C_2$ & $C_3$ & $C_4$ & $C_5$ \\
$|\phi_1\!\!>=|0000100000111111\!\!>$ & $0.107 \pm 0.009$& $0.0081 \pm 0.0023$& $0.092 \pm 0.017$& $0.0048 \pm 0.0001$& $0.0054 \pm 0.0003$\\
$|\phi_2\!\!>=|0010100000011111\!\!>$ & $0.100 \pm 0.003$& $0.0103 \pm 0.0011$& $0.069 \pm 0.007$& $0.0068 \pm 0.0001$& $0.0099 \pm 0.0003$\\
$|\phi_3\!\!>=|0000011000110111\!\!>$ & $0.110 \pm 0.005$& $0.0130 \pm 0.0016$& $0.080 \pm 0.012$& $0.0076 \pm 0.0001$& $0.0098 \pm 0.0003$\\
$|\phi_4\!\!>=|1000100011001011\!\!>$ & $0.103 \pm 0.003$& $0.0210 \pm 0.0007$& $0.018 \pm 0.005$& $0.0094 \pm 0.0001$& $0.0140 \pm 0.0002$\\
\end{tabular}

%% file: sinaioscarxiv.bbl
\begin{thebibliography}{10}

\bibitem{loschmidt} Loschmidt J., 
        {\it \"Uber den Zustand des Wärmegleichgewichts eines Systems von 
         K\"orpern mit Rücksicht auf die Schwerkraft}, 
         Sitzungsberichte der Akademie der Wissenschaften, Wien, 
         {\bf II-73} (1876) 128-142.
\bibitem{boltzmann} Boltzmann L., 
        {\it \"Uber die Beziehung eines allgemeine mechanischen Satzes zum zweiten 
          Haupsatze der W\"rmetheorie}, 
         Sitzungsberichte der Akademie der Wissenschaften, Wien, {\bf II-75} (1877) 67-73.
\bibitem{mayer} Mayer J.E. and Goeppert-Mayer M.
                {\it Statistical mechanics}, Wiley, New York (1977).
\bibitem{jalabert} Gousev A., Jalabert R.A., Pastawski H.M. and Wisniacki D.A.,
         {\it Loschmidt echo},
         Scholarpedia, {\bf 7(8)} (2012) 11687.
\bibitem{arnold} Arnold V. and Avez A.,
         {\it Ergodic [roblems in classical mechanics}
          Benjamin, New York (1968).
\bibitem{sinaibook} Cornfeld I.P., Fomin S.V. and Sinai Ya.G.,
         {\it Ergodic theory},
          Springer-Verlag, New York (1982).
\bibitem{chirikov} Chirikov B.V.
         {\it A universal instability of many-dimensional oscillator systems},
         Physics Reports {\bf 52} (1979) 263.
\bibitem{lichtenberg} Lichtenberg A. and Lieberman M.,
          {\it Regular and chaotic dynamics},
          Springer, New York (1992).
\bibitem{sinaibil} Sinai Ya.G.,
          {\it Dynamical  systems  with  elastic  reflections. 
           Ergodic  properties  of  dispersing  billiards},
           Uspekhi  Mat. Nauk {\bf 25(2)} (1970) 141
            [English trans.:  Russian Math. Surveys {\bf 25(2)} (1970) 137].
\bibitem{gutzwiller} Gutzwiller M.C.,
           {\it Chaos in classical and quantum mechanics},
            Springer, New York (1990).
\bibitem{haake} Haake F.,
            {\it Quantum signatures of chaos},
             Springer, Berlin (2010).
\bibitem{stockmann} Stockmann H.-J.,
            {\it Microwave billiards and quantum chaos},
            Scholarpedia {\bf 5(10)} (2010) 10243.
\bibitem{bohigas1984} Bohigas O., Giannoni M.J. and Schmit C.,
         {\it Characterization of chaotic quantum spectra
          and universality of level fluctuation},
         Phys. Rev. Lett. {\bf 52} (1984) 1.
\bibitem{wigner} Wigner E., 
        {\it Random matrices in physics},
        SIAM Rev. {\bf 9(1)} (1967) 1.
\bibitem{mehta} Mehta M.L.,
        {\it Random matrices},
        Elsvier Academic Press, Amsterdam (2004).
\bibitem{ullmo} Ullmo D.,
         {\it Bohigas-Giannoni-Schmit conjecture},
          Scholarpedia  {\bf 11(9)} (2016) 31721.
\bibitem{chirikov1981} Chirikov B.V, Izrailev F.M. and Shepelyansky D.L.,
         {\it Dynamical stochasticity in classical and quantum mechanics},
          Sov. Scient. Rev. (Harwood Acad. Publ., Chur, Switzerland) 
          {\bf 2C} (1981) 209.
\bibitem{fishman} Fishman S., Grempel D.R. and Prange R.E,
          {\it Chaos, quantum recurrences, and Anderson localization},
           Phys. Rev. Lett. {\bf 49} (1982) 509.
\bibitem{chirikov1988} Chirikov B.V, Izrailev F.M. and Shepelyansky D.L.,
         {\it Quantum chaos: localization vs. ergodicity},
         Physica D {\bf 33} (1988) 77.
\bibitem{fishmanschol} Fishman S., 
         {\it Anderson localization and quantum chaos maps},
         Scholarpedia {\bf 5(8)} (2010) 9816.
\bibitem{anderson} Anderson P.W.,
         {\it Absence of diffusion in certain random lattices},
          Phys. Rev. {\bf 109} (1958) 1492.
\bibitem{rough1} Frahm K.M. and Shepelyansky D.L., 
         {\it Quantum localization in rough billiards}, 
         Phys. Rev. Lett. {\bf 78} (1997) 1440.
\bibitem{rough2}  Frahm K.M. and Shepelyansky D.L.,
         {\it Emergence of quantum ergodicity in rough billiards},
          Phys. Rev. Lett. {\bf 79} (1997) 1833.
\bibitem{bohrcorres} Bohr N.,
          {\it \"Uber die Serienspektra der Element},
             Zeitschrift f\"ur Physik  {\bf 2(5)} (1920) 423.
\bibitem{ehrenfest} Ehrenfest P., 
         {\it Bemerkung \"uber die angen\"aherte G\"ultigkeit der klassischen Mechanik 
         innerhalb der Quantenmechanik},
         Zeitschrift f\"ur Physik {\bf 45 (7–8)} (1927) 455.
\bibitem{dls1981} Shepelyanskii D.L., 
         {\it Dynamical stochasticity in nonlinear quantum systems},
          Theor. Math. Phys. {\bf 49(1)} (1981) 925.
\bibitem{dls1983} Shepelyansky D.L.,
          {\it  Some statistical properties of simple classically stochastic quantum systems},
           Physica D {\bf 8} (1983) 208.
\bibitem{stmapscholar} Chirikov B. and Shepelyansky D., 
          {\it Chirikov standard map}, Scholarpedia {\bf 3(3)} (2008) 3550.
\bibitem{bohr} Bohr A. and Mottelson B.R.,
         {\it Nuclear structure},
          Benjamin, New York {\bf 1} (1969) 284.
\bibitem{guhr} Guhr T., Muller-Groeling A. and Weidenmuller H.A.,
        {\it Random-matrix theories in quantum physics: common concepts},
        Phys. Rep. {\bf 299} (1998) 189.
\bibitem{french1} French J.B., and Wong S.S.M.,
         {\it Validity of random matrix theories for many-particle systems},
         Phys. Lett. B {\bf 33} (1970) 449.
\bibitem{bohigas1} Bohigas O. and Flores J.,
        {\it Two-body random Hamiltonian and level density},
         Phys. Lett. B {\bf 34} (1971) 261.
\bibitem{french2} French J.B. and Wong S.S.M.,
         {\it Some random-matrix level and spacing distributions for
         fixed-particle-rank interactions},
         Phys. Lett. B {\bf 35} (1971) 5.
\bibitem{bohigas2} Bohigas O. and Flores J.,
        {\it Spacing and individual eigenvalue distributions of two-body
         random Hamiltonians},
         Phys. Lett. B {\bf 35} (1971) 383.
\bibitem{thouless} Thouless D.J.,
          {\it Maximum Metallic Resistance in Thin Wires},
           Phys. Rev. Lett. {\bf 39} (1977) 1167.
\bibitem{imry} Imry Y.,
           {\it Introduction to mesoscopic physics},
           Oxford University Press, Oxford (2002).
\bibitem{akkermans} Akkermans E. and Montambaux G.,
          {\it Mesoscopic physics of electrons and photons},
           Cambridge Univ. Press, Cambridge (2007).
\bibitem{aberg1} {\AA}berg S.,
         {\it Onset of chaos in rapidly rotating nuclei}, 
         Phys. Rev. Lett. {\bf 64} (1990) 3119.
\bibitem{aberg2} {\AA}berg S.,
         {\it Quantum chaos and rotational damping},
         Prog. Part. Nucl. Phys. {\bf 28} (1992) 11.
\bibitem{dlsnobel} Shepelyansky D.L.,
         {\it Quantum chaos and quantum computers},
         Physica Scripta {\bf T90} (2001) 112.
\bibitem{jacquod} Jacquod P. and Shepelyansky D.L.,
         {\it Emergence of quantum chaos in finite interacting Fermi systems},
         Phys. Rev. Lett. {\bf 79} (1997) 1837.
\bibitem{sushkov} Shepelyansky D.L. and Sushkov O.P.,
         {\it Few interacting particles in a random potential},
         Europhys. Lett. {\bf 37} (1997) 121.
\bibitem{mirlin1} Gornyi I.V., Mirlin A.D.  and Polyakov D.G.,
        {\it Many-body delocalization transition and relaxation in a quantum dot},
         Phys. Rev. B {\bf 93} (2016) 125419.
\bibitem{mirlin2}  Gornyi I.V., Mirlin A.D., Polyakov D.G.  and Burin A.L.,
         {\it Spectral diffusion and scaling of many-body delocalization transitions},
         Ann. Phys. (Berlin) {\bf 529} (2017) 1600360.
\bibitem{kolovsky2017} Kolovsky A.R. and Shepelyansky D.L.,
        {\it Dynamical thermalization in isolated quantum dots and black holes},
        EPL {\bf 117} (2019) 10003.
\bibitem{frahmtbrim} Frahm K.M. and Shepelyansky D.L., 
         {\it Dynamical decoherence of 
         a qubit coupled to a quantum dot or the SYK black hole},
         Eur. Phys. J. B {\bf 91} (2018) 257.
\bibitem{landau} Landau L.D. and Lifshitz E.M.,
         {\it Statistical mechanics},
         Wiley, New York (1976).
\bibitem{ermannsinai} Ermann L., Vergini E. and Shepelyansky D.L.,
         {\it Dynamics and thermalization a Bose-Einstein condensate in a Sinai-oscillator trap},
         Phys. Rev. A {\bf 94} (2016) 013618.
\bibitem{ketterle1} Davis K.B., Mewes M.-O., Andrews M.R.,
         van Druten N.J., Durfee D.S., Kurn D.M., and Ketterle W.,
         {\it Bose-Einstein Condensation in a Gas of Sodium Atoms}, 
         Phys. Rev. Lett. {\bf 75} (2015) 3969.
\bibitem{ketterle2} Anglin J.A. and  Ketterle W.,
         {\it Bose–Einstein condensation of atomic gases},
          Nature {\bf 416} (2002)  211.
\bibitem{ketterle3} Ketterle W.,
         {\it Nobel lecture: When atoms behave as waves:
         Bose-Einstein condensation and the atom laser},
         Rev. Mod. Phys. {\bf 74} (2002) 1131.
\bibitem{roati1} Valtolina G., Scazza F., Amico A., Burchianti A., Recati A., 
          Enss T., Inguscio M.,
         Zaccanti M.  and Roati G.,
         {\it Exploring the ferromagnetic behaviour of a repulsive 
           Fermi gas through spin dynamics},
          Nature Phys. {\bf 13} (2017) 704.
\bibitem{roati2} Burchianti A., Scazza F.,Amico A.,  Valtolina G.,  Seman J.A.,  Fort C., 
          Zaccanti M., Inguscio M. and Roati G.,
         {\it Connecting dissipation and phase slips in a Josephson junction
          between fermionic superfluids},
         Phys. Rev. Lett. {\bf 120} (2018) 025302.
\bibitem{sachdevprl} Sachdev S. and Ye J.,
         {\it Gapless spin-fluid ground state in a random quantum Heisenberg magnet},
         Phys. Rev. Lett. {\bf 70} (1993) 3339.
\bibitem{kitaev} Kitaev A., 
         {\it A simple model of quantum holography},
          Video talks at KITP Santa Barbara, April 7 and May 27 (2015).
\bibitem{sachdevprx} Sachdev S.,
          {\it Bekenstein-Hawking entropy and strange metals},
          Phys. Rev. X {\bf 5} (2015) 041025.
\bibitem{rosenhaus} Polchinski J. and Rosenhaus V., 
         {\it The spectrum in the Sachdev-Ye-Kitaev model},
         JHEP {\bf 04} (2016) 1.
\bibitem{maldacena} Maldacena J. and Stanford D.,
         {\it Remarks on the Sachdev-Ye-Kitaev model},
          Phys. Rev. D {\bf 94} (2016) 106002.
\bibitem{garcia1} Garcia-Garcia A.M. and Verbaarschot J.J.M.,
         {\it Spectral and thermodynamic properties of the Sachdev-Ye-Kitaev model},
         Phys. Rev. D {\bf 94} (2016) 126010.
\bibitem{huse} Nandkishore R. and Huse D.A.,
         {\it Many-body localization and thermalization in quantum statistical mechanics},
         Annu. Rev. Condens. Matter Phys. {\bf 6} (2015) 15.
\bibitem{polkovnikov} Alessiom L.D., Kafri Y., Polkovnikov A. and Rigol M.,
         {\it From quantum chaos and eigenstate thermalization 
         to statistical mechanics and thermodynamics},
         Adv. Phys. {\bf 65} (2016) 239.
\bibitem{borgonovi} Borgonovi F., Izrailev F.M., Santos L.F.  and Zelevinsky V.G.,
         {\it  Quantum chaos and thermalization in isolated systems of interacting particles},
         Phys. Rep. {\bf 626} (2016) 1.
\bibitem{alet} Alet F. and Laflorencie N.,
         {\it Many-body localization: an introduction and selected topics},
         Comptes Rendus Physique {\bf 19} (2018) 498.
\bibitem{gribakin1}  Gribakin G.F. and Flambaum V.V.,
         {\it Calculation of the scattering length in atomic collisions
          using the semiclassical approximation},
          Phys. Rev. A {\bf 48} (1999) 1998.
\bibitem{gribakin2} Flambaum V.V., Gribakin G.F. and Harabati C.,
         {\it Analytical calculation of cold-atom scattering},
         Phys. Rev. A {\bf 59} (1993) 546.
\bibitem{wilkens} Busch T., Englert B.-G., Rzazewski K. and Wilkens M.,
          {\it Two cold atoms in a harmonic trap},
          Foundations Phys. {\bf 28} (1998) 549.
\bibitem{kohler} Kohler T., Goral K. and Julienne P.,
         {\it Production of cold molecules via magnetically tunable Feshbach resonances},
         Rev. Mod. Phys. {\bf 78} (2006) 1311.
\bibitem{flambaum} Flambaum V.V. and Izrailev F.M.,
         {\it Distribution of occupation numbers in finite Fermi systems and 
         role of interaction in chaos and thermalization},
         Phys. Rev. E {\bf 55} (1997) R13(R).
\bibitem{quantware} Web page of quantware with further data: videos, 
           figures etc. 
\bibitem{ketterlefermi} Ketterle, W., Durfee D.S., and Stamper-Kurn D.M.,
            {\it Making, probing and understanding Bose-Einstein condensates},
             in Proceedings  of  the  International  School  of  Physics  
            ‘‘Enrico Fermi,’’ Course CXL, Eds. Inguscio M., Stringari S. and
             Wieman C.E., IOS Press, Amsterdam, p.67 (1999);
             arXiv:cond-mat/9904034v2 (1999).
\bibitem{tip} Shepelyansky D.L.,
            {\it Coherent propagation of two interacting particles in a random potential},
            Phys. Rev. Lett. {\bf 73} (1994) 2607.
\bibitem{hoogerland} Ullah A. and Hoogerland M.D., 
           {\it Experimental observation of Loschmidt time reversal of a quantum chaotic system},             Phys. Rev. E 83: 046218 (2012).
\bibitem{poincare} Poincare H.,
         {\it Sur les equations de la dynamique et le probleme des trois corps},
         Acta Mathematica {\bf 13} (1890) 1.


\end{thebibliography}
